\documentclass[final,1p,times,twocolumn]{elsarticle}

\usepackage[utf8]{inputenc}
\usepackage{amssymb}
\usepackage{amsthm}
\usepackage{amsmath}
\usepackage{lineno}
\usepackage{subfigure} 
\usepackage[colorlinks=true,linkcolor=black, citecolor=blue, urlcolor=blue]{hyperref}

\usepackage{booktabs}
\usepackage{multirow}
\usepackage{rotating}
\usepackage{booktabs}
\usepackage{url}

\biboptions{numbers,sort&compress}
\usepackage{textcomp}

\begin{document}
\begin{frontmatter}

\title{Electromagnetic Showers Beyond Shower Shapes}

\author[a,c]{Luke de Oliveira}
\ead{lukedeo@ldo.io}
\author[a]{Benjamin Nachman}
\ead{bnachman@cern.ch}
\author[a,b]{Michela Paganini}
\ead{michela.paganini@yale.edu}

\address[a]{Lawrence Berkeley National Laboratory, 1 Cyclotron Road, Berkeley, CA 94720, USA}
\address[b]{Yale University, Department of Physics, 217 Prospect Street, New Haven, CT 06511, USA}
\address[c]{Vai Technologies, 350 Rhode Island Street, San Francisco, CA 94103, USA}

\begin{abstract}
Correctly identifying the nature and properties of outgoing particles from high energy collisions at the Large Hadron Collider is a crucial task for all aspects of data analysis. Classical calorimeter-based classification techniques rely on shower shapes -- observables that summarize the structure of the particle cascade that forms as the original particle propagates through the layers of material. This work compares shower shape-based methods with computer vision techniques that take advantage of lower level detector information. In a simplified calorimeter geometry, our DenseNet-based architecture matches or outperforms other methods on $e^+$-$\gamma$ and $e^+$-$\pi^+$ classification tasks. In addition, we demonstrate that key kinematic properties can be inferred directly from the shower representation in image format. 
\end{abstract}

\begin{keyword}
Deep Learning \sep Classification \sep Large Hadron Collider \sep Calorimetry \sep Electromagnetic Showers 
\end{keyword}

\end{frontmatter}

\section{Introduction}
Treating calorimeters as digital cameras has had a long history in high energy particle physics~\cite{Thomson:302717,Bock:323781}.  Calorimeter cells are like the pixels in a camera and the energy deposited is like the pixel intensity.  Recently, deep neural networks have revolutionized image processing, with significant improvement over traditional techniques on a variety of tasks.  Many of these modern techniques have already been applied to high energy physics in the context of \textit{jet images}~\cite{Cogan:2014oua} for classification~\cite{deOliveira:2015xxd,Baldi:2016fql,Barnard:2016qma,Almeida:2015jua,Kasieczka:2017nvn,komiske2017,CMS-DP-2017-027,ATL-PHYS-PUB-2017-017, Macaluso:2018tck, Fraser:2018ieu, Guo:2018hbv,Choi:2018dag,Komiske:2018oaa}, regression~\cite{Komiske:2017ubm}, and generation~\cite{lagan,Musella:2018rdi} as well as in the context of neutrino identification and classification~\cite{Racah:2016gnm,Aurisano:2016jvx,Acciarri:2016ryt,Renner:2016trj, Ai:2018juq} in liquid argon time projection chambers and generation for cosmic ray detector simulations~\cite{Erdmann:2018kuh}.  Jet image generation has recently been extended to particle showers in a longitudinally segmented calorimeter~\cite{Paganini:2017hrr,Paganini:2017dwg, deOliveira:2017rwa} using Generative Adversarial Networks (GANs).  
In this paper, we explore classification and regression for particles in a longitudinally segmented calorimeter using the techniques developed in Ref.~\cite{deOliveira:2017rwa} as well as other ideas from machine learning.

Traditional techniques for identifying particles in a longitudinally segmented calorimeter rely on a small number of \textit{shower shapes}.  These moments of the three-dimensional shower profile are powerful tools for identifying and calibrating photons and electrons~\cite{Aaboud:2016vfy,Aaboud:2016yuq,Aad:2014nim,Aad:2014fxa} as well as extracting pointing information for photons~
\cite{Aad:2012tfa,Aad:2014gfa}. Our goal is to show how much one can gain from using deep learning techniques, treating the calorimeter region around a particle shower as a series of digital images. Unlike a typical RGB image, longitudinally segmented calorimeter images are sparse, without smooth features or sharp edges, and have a causal relationship between layers, so it is not sufficient to treat each layer independently. For these reasons, image processing techniques must be adapted to fit this use case. Due to the great promise of deep learning-based computer vision tools, related efforts exist within the high energy physics community using similar data sets~\cite{gan4alice,cambridge,lcd,geantv,nyu, dampe}. We provide a set of baselines to help reduce the search space towards optimal solutions.

Section~\ref{sec:setup} introduces the simulation setup and the data set. 
Section~\ref{sec:classification} describes the machine learning methods tested on the two classification tasks, and provides experimental results. Section~\ref{sec:regression} focuses on the regression task and its outcome. The paper concludes in Sec.~\ref{sec:conclusion} with future outlook.

\section{Experimental Setup and Data}
\label{sec:setup}
The public data set~\cite{dataset} used for training is composed of 500,000 $e^+$, 500,000 $\pi^+$, and 400,000 $\gamma$ showers induced by the electromagnetic and nuclear interactions that the incident and secondary particles undergo as they propagate through the electromagnetic calorimeter.  The geometry of the detector, built from a modified version of the \textsc{Geant4} B4 example~\cite{b4example}, consists of a cubic section (volume $480$ mm$^3$) along the radial ($z$) direction of an ATLAS-inspired electromagnetic calorimeter, at a distance of $z_0 = 288$ mm from the origin. The volume is segmented along its radial dimension into three layers of depth 90 mm, 347 mm, and 43 mm, each composed by flat alternating layers of lead (absorber) and LAr (active material) of thickness 2 mm and 4 mm, respectively. Each of the three sub-volumes has a different resolution, with voxels of dimensions summarized in Table~\ref{table:detector}. The energy in each layer is the sum of over both the active and inactive sub-layers.  In practice, the energy deposited in the absorber is not measured, but this complication is not part of the dataset~\cite{dataset}. As a result of the simplifications used in constructing the dataset, only relative gains are important and absolute performance should not be compared with results from current experimental publications. 

\begin{table}[]
\centering
\caption{Specifications of the calorimeter layers}
\label{table:detector}
\begin{tabular}{cccccc}
\toprule
Layer Number           & \begin{tabular}[c]{@{}c@{}}Depth in $z$\\direction (mm)\end{tabular} & $N_{\mathrm{cells}, x}$ & \begin{tabular}[c]{@{}c@{}}Width of cell in $x$\\direction (mm)\end{tabular} & $N_{\mathrm{cells}, y}$ & \begin{tabular}[c]{@{}c@{}}Width of cell in $y$\\direction (mm)\end{tabular} \\
\midrule
0    &  90  &  3  & 160  & 96  & 5  \\
1    & 347  & 12  & 40   & 12  & 40 \\
2    &  43  & 12  & 40   & 6   & 80 \\   
\bottomrule
\end{tabular}
\end{table}

In the native \textsc{Geant4} coordinate system (G), the $z$ direction corresponds to the radial direction of a collider experiment (C). The following coordinate transformations relate the two coordinate systems:

\begin{equation}
\hat{x}_{G} = \hat{y}_{C}; \ \ \ \ \ \hat{y}_{G} = \hat{z}_{C}; \ \ \ \ \ \hat{z}_{G} = \hat{x}_{C}
\label{eq:transform}
\end{equation}
where the subscripts identify the coordinate system, and $\hat{z}_{A}$ is along the beam line, $\hat{x}_{C}$ points towards the center of the LHC ring, and $\hat{y}_{C}$ points upwards towards the sky. To make the results most easily interpretable, collider coordinates will be used when presenting the regression results. In the dataset, the incoming particles are shot in a cone around the $\hat{z}_{G}$ with maximum angle of incidence of $5^\circ$ and maximum displacement from the origin of 5 cm. The detector volume is big enough to contain more than 99\% of all showers in the training and test sets. More information about the coordinate systems and the particle distributions in this data set are available in ~\ref{appA}.

\section{$e^+$-$\gamma$ and $e^+$-$\pi^+$ Classification}
~\label{sec:classification}

Electrons and photons are mostly stopped by our calorimeter while charged pions leave only a fraction of their energy in the three layers.  The fluctuations in the shower are narrower and more Gaussian for electrons and photons compared to charged pions.  Photon showers are slightly deeper than electron ones because the mean free path for pair production is slightly longer than the distance required for electrons to loose a significant fraction of their energy.  For a brief review of the different calorimeter signatures of various particles, see e.g.~\cite{pdg}.

\subsection{Method}
\label{sec:method}
Different machine learning methods~\footnote{Code is available at \texttt{\url{https://github.com/hep-lbdl/CaloID}}} are tested on two classification problems ($e^+$ versus $\gamma$, and $e^+$ versus $\pi^+$). The scope of this approach is to document both successful and unsuccessful attempts, and to inform the community on what techniques appear to be more promising and worth pursuing.

All neural networks are built using \textsc{Keras}~\cite{keras} with \textsc{TensorFlow}~\cite{tensorflow} backend, and trained on an NVIDIA GeForce$^\circledR$ GTX TITAN Xp with the Adam~\cite{adam} optimizer to minimize the cross-entropy between the predicted and target distributions. The learning rate is set to 0.001 for all networks, except for 3 three-stream classifiers described below in the $e^+ $-$ \pi^+$ task, where a quick hyper-parameter scan found a value of 0.0001 to result in higher, more stable performance.  The following sections describe the methods presented in the results (Sec.~\ref{sec:results}).

\subsubsection{Fully-Connected Network on Shower Shapes}
\label{ssec:ss}
As a first baseline, we train a simple six-layer feed-forward neural network to learn a discriminant from 20 \textit{shower shape} input variables described in Table~\ref{table:ss} (same as those used in Ref.~\cite{Paganini:2017hrr,Paganini:2017dwg}).  We adopt a simple neural network architecture with five fully connected - LeakyReLU~\cite{leaky} - dropout~\cite{dropout} - batch normalization~\cite{batchnorm} blocks, with hidden representations of size 512, 1024, 2048, 1024, and 128 respectively, and a final one-dimensional output with sigmoid activation. The network has a total of 4,873,985 trainable parameters, and the batch size is chosen to be 128.

\renewcommand{\arraystretch}{1.1}
\begin{table}
\caption{Description and mathematical definition of the 1-dimensional shower shape observables, defined as functions of $\mathcal{I}_i$, the vector of pixel intensities for an image in layer $i$, where $i \in \{0, 1, 2\}$}
\centering
\begin{tabular}{ccp{4.0cm}}
    \toprule
    \textbf{Shower Shape Variable} & \textbf{Formula} & \textbf{Notes}
    \\
    \midrule
    \addlinespace[1em]
    $E_i$ & $E_i = \sum_\mathrm{pixels} \mathcal{I}_i $ & Energy deposited in the $i^{th}$ layer of calorimeter\\
    \addlinespace[1em]
    $E_\mathrm{tot}$ & $E_\mathrm{tot}=\sum\limits_{i=0}^2 E_i$ & Total energy deposited in the electromagnetic calorimeter\\
    \addlinespace[1em]
    $f_i$ & $f_i = E_i / E_\mathrm{tot}$ & Fraction of measured energy deposited in the $i^{th}$ layer of calorimeter\\
    \addlinespace[1em]
     $E_{\mathrm{ratio}, i}$ & $
    \dfrac{\mathcal{I}_{i,(1)} - \mathcal{I}_{i,(2)}}{\mathcal{I}_{i,(1)} + \mathcal{I}_{i,(2)}}
    $ & Difference in energy between the highest and second highest energy deposit in the cells of the $i^{th}$ layer, divided by the sum\\
    \addlinespace[1em]
    $d$ & $d = \max\{i: \max(\mathcal{I}_i) > 0\}$ & Deepest calorimeter layer that registers non-zero energy\\
    \addlinespace[1em]
    Depth-weighted total energy, $l_d$ & $l_d=\sum\limits_{i=0}^2 i\cdot  E_i$ & The sum of the energy per layer, weighted by layer number \\
    \addlinespace[1em]
    Shower Depth, $s_d$ & $s_d = l_d / E_\mathrm{tot}$ & The energy-weighted depth in units of layer number \\
    \addlinespace[1em]
    Shower Depth Width, $\sigma_{s_d}$ & $\sigma_{s_d} = \sqrt{\frac{\sum\limits_{i=0}^2 i^2\cdot  \mathcal{I}_i}{E_\mathrm{tot}} - \left(\frac{\sum\limits_{i=0}^2 i\cdot \mathcal{I}_i}{E_\mathrm{tot}}\right)^2}$ &The standard deviation of $s_d$ in units of layer number \\
    \addlinespace[1em]
    $i^{\mathrm{th}}$ Layer Lateral Width, $\sigma_i$& $\sigma_i =\sqrt{\frac{\mathcal{I}_i \odot H^2}{E_i} - \left(\frac{\mathcal{I}_i \odot H}{E_i}\right)^2}$& The standard deviation of the transverse energy profile per layer, in units of cell numbers \\
    \addlinespace[1em]
    Sparsity$_i$ & $\frac{\sum_\mathrm{pixels} \mathcal{I}_i > 0}{N_{\mathrm{pixels},i}}$ & Percentage of non-zero pixels in a layer \\
    \bottomrule
\end{tabular}
\label{table:ss}
\end{table}

\subsubsection{Fully-Connected Network on Individual Pixel Intensities}
\label{ssec:fcn_pixels}
The network structure is identical to the one described in Sec.~\ref{ssec:ss}, except for the first layer that now receives as inputs the 504 calorimeter pixel intensities from a shower representation, as opposed to the 20 shower shape variables used in the previous benchmark. The network now has a total of 5,121,793 trainable parameters, and the batch size is chosen to be 128.
    
\subsubsection{Three-Stream Locally-Connected Network}
\label{ssec:lagan}
Locally-connected layers have shown promising results compared to their convolutional counterpart in both classification and generation tasks~\cite{lagan} on high energy physics datasets, where domain-specific preprocessing techniques allow to rotate, crop, and center images with very high sparsity, dynamic range, and physical meaning associate to pixel intensities~\cite{Cogan:2014oua}. Unlike the case of natural images, this application domain has been shown to benefit from the location specificity of filters learned by locally-connected layers. 

These were recently employed in the design of both the generator and the discriminator networks in Location-Aware Generative Adversarial Networks (LAGAN)~\cite{lagan}, and tested in their multi-stream evolution (CaloGAN)~\cite{Paganini:2017hrr}. We draw inspiration from previous applications in generative modeling to test a similar design for the classification tasks presented in this work.

The network consists of three streams of LAGAN-style~\cite{lagan} blocks, each aimed at processing images from one of the three calorimeter layers, and each containing a convolutional layer and three sets of locally-connected layers, batch normalization, and leaky rectified linear units. The features learned from the three streams provide different representations of individual showers, and are then concatenated and processed through a top fully-connected layer with a sigmoid activation. The network has a total of 17,525,697 trainable parameters, and the batch size is chosen to be 128. 

\subsubsection{Three-Stream Convolutional Network}
\label{ssec:3scnn}
Although locally-connected layers were empirically found to work well with jet images centered at the origin~\cite{lagan_data}, the advantage of using them over convolutional layers is expected to fade away as showers are produced at different incoming angles and positions. In fact, convolutional layers are designed to exploit feature translation invariance.

The architecture in Sec.~\ref{ssec:lagan} is modified by replacing all locally-connected layers with equivalent convolutional layers. The new network has a total of 7,434,881 trainable parameters, and the batch size is chosen to be 128.

\subsubsection{Three-Stream DenseNet}
\label{ssec:3sdnet}
Densely Connected Convolutional Networks (DenseNets)~\cite{densenet} were introduced as an elegant solution to maximize information flow by reducing the path from input to output, in order to counter the vanishing gradient problem in very deep convolutional networks. DenseNets devise connections such that every layer receives as inputs the concatenated feature maps from every previous layer, and contributes its feature maps to every subsequent layer. These redundant connections favor feature reuse and persistence, to the point that the last classification layer will have at its disposal all of the features built by all previous layers in the network, therefore gaining access to different levels of feature representation. 

The network is, by design, very parameter efficient, with only 351,057 trainable parameters. To match the other benchmarks, the batch size is set to 128. 

\subsection{Experimental Results}
\label{sec:results}

\begin{figure}[h!]
\centering
\subfigure[ROC curves for $e^+ $-$ \gamma$ classification]{
\centering
        \includegraphics[width=0.45\textwidth]{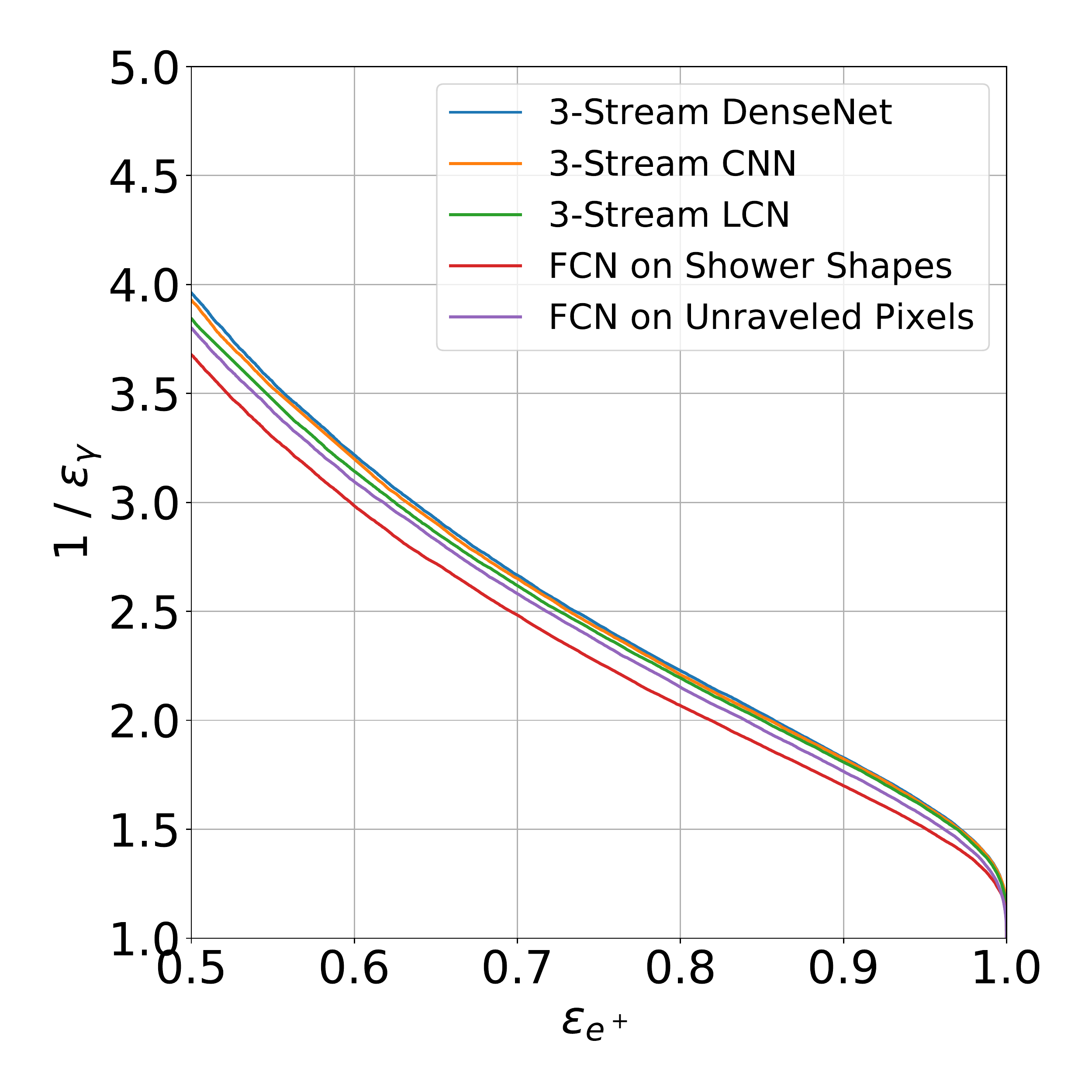}
        \label{fig:rocsa}
}%
\subfigure[ROC curves for $e^+ $-$ \pi^+$ classification]{
\centering
        \includegraphics[width=0.45\textwidth]{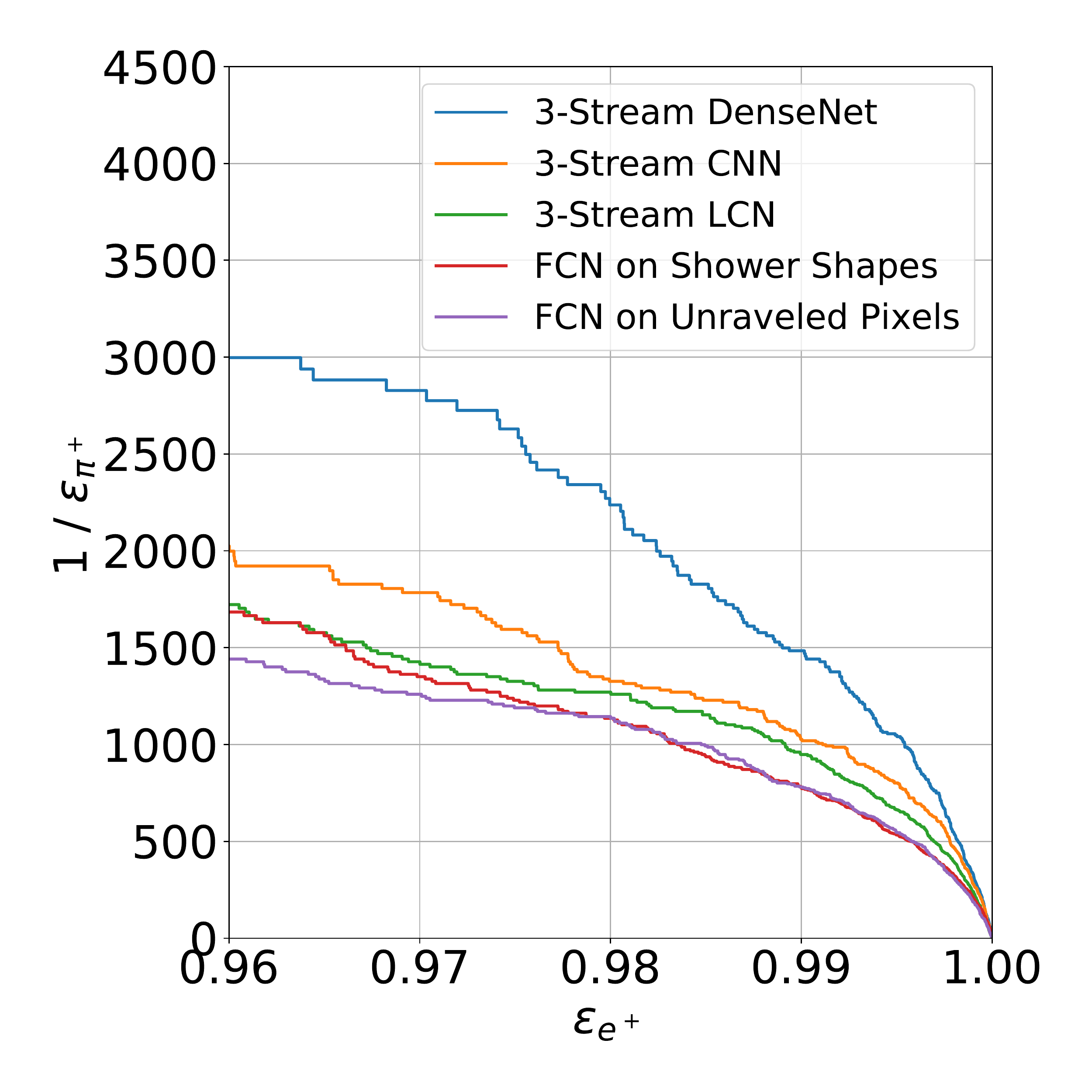}
        \label{fig:rocsb}
}
\caption{These performance plots illustrate the trade-off between maximizing the true positive ratio for positron identification (on the $x$-axis) and maximizing the background rejection, the inverse of the false positive ratio (on the $y$-axis). The five curves represent the performance of the following classifiers: in blue, the three-stream DenseNet~\ref{ssec:3sdnet}; in orange, the three-stream convolutional network~\ref{ssec:3scnn}; in green, the three-stream locally-connected network~\ref{ssec:lagan}; in red, the fully-connected network on shower shapes~\ref{ssec:ss}; in purple, the fully-connected network on individual pixel intensities~\ref{ssec:fcn_pixels}.}
\label{fig:rocs}
\end{figure}

We examine the performance of the binary classifiers described in Sec.~\ref{sec:method} using receiver operating characteristic (ROC) curves (Fig.~\ref{fig:rocs}). The different efficiency ranges depicted on the axes of Figures~\ref{fig:rocsa} and~\ref{fig:rocsb} illustrate the difference in complexity between the two tasks: while charged pions are easier to separate from positrons and only the high signal efficiency range is displayed, photons share similar signatures in the electromagnetic calorimeter compared to positrons, yielding worse overall background rejection.

In both classification tasks, the DenseNet outperforms all other architectures and does so with an order of magnitude fewer parameters (or two, in the case of the less parameter-efficient locally-connected setup).

In the harder $e^+$ versus $\gamma$ scenario, the relative performance differentials with respect to the shower shapes-based classifier are provided, for five different $e^+$ efficiency points, in Table~\ref{table:egamma}. Similar results are provided in Table~\ref{table:epi} for the $e^+$ versus $\pi^+$ classification task.

\begin{table}[]
\centering
\caption{Percentage relative increase or decrease in $\gamma$ rejection at five different $e^+$ efficiency working points compared to the baseline fully-connected network trained on shower shape variables}
\label{table:egamma}
\begin{tabular}{lllllll}
\toprule
& & \multicolumn{5}{c}{$e^+$ efficiency}\\
\cmidrule{3-7}
&  & 60\%   & 70\%   & 80\%   & 90\%   & 99\%   \\
\midrule
\multirow{5}[0]{*}{\begin{sideways}Model\end{sideways}}
& FCN on shower shapes
    & -      & -      & -      & -      & - \\
& FCN on unraveled pixels                        & +3.7\% & +3.9\% & +4.1\% & +3.9\% & +2.1\% \\
& 3-Stream Locally-Connected                     & +5.3\% & +5.4\% & +6.1\% & +6.4\% & +5.2\% \\
& 3-Stream Conv Net                              & +5.5\% & +6.8\% & +6.9\% & +7.3\% & +6.0\% \\
& 3-Stream DenseNet                     & \textbf{+7.5\%} & \textbf{+7.4\%} & \textbf{+7.7\%} & \textbf{+7.6\%} & \textbf{+6.4\%} \\
\bottomrule
\end{tabular}

\centering
\caption{Percentage relative increase or decrease in $\pi^+$ rejection at five different $e^+$ efficiency working points compared to the baseline fully-connected network trained on shower shape variables}
\label{table:epi}
\begin{tabular}{lllllll}
\toprule
& & \multicolumn{5}{c}{$e^+$ efficiency}\\
\cmidrule{3-7}
&  & 96\%   & 97\%   & 98\%   & 99\%  & 99.99\%  \\
\midrule
\multirow{5}[0]{*}{\begin{sideways}Model\end{sideways}}
& FCN on shower shapes
    & -      & -      & -      & -      & - \\
& FCN on unraveled pixels                       & --14.4\% & --7.6\% & +0.76\% & +0.0\% & --34.6\%\\
& 3-Stream Locally-Connected                    & +2.3\% & +4.8\% & +11.9\% & +22.3\% & --43.7\%\\
& 3-Stream Conv Net                             & +20.3\% & +31.0\% & +17.9\% & +32.4\% & --6.8\%\\
& 3-Stream DenseNet                             & \textbf{+81.6\%} & \textbf{+107.5\%} & \textbf{+100.0\%} & \textbf{+90.1\%} & \textbf{+34.9\%}\\
\bottomrule
\end{tabular}
\end{table}

It is useful to visualize the output of the DenseNet in comparison to the output of the baseline shower shapes-based classifier, in order to identify critical regions in which the two classifiers are not in agreement on the shower labels. The following analysis refers to the $e^+$-$\gamma$ classification task, and will try to identify shortcomings of a shower shapes-based method. 

\begin{figure}
\centering
\subfigure[2D distribution of output scores for $e^+$ showers (target: 0). The box highlight a subset of showers that are correctly classified by the DenseNet and incorrectly classified by the shower shapes-based network.]{
\centering
        \includegraphics[width=0.45\textwidth]{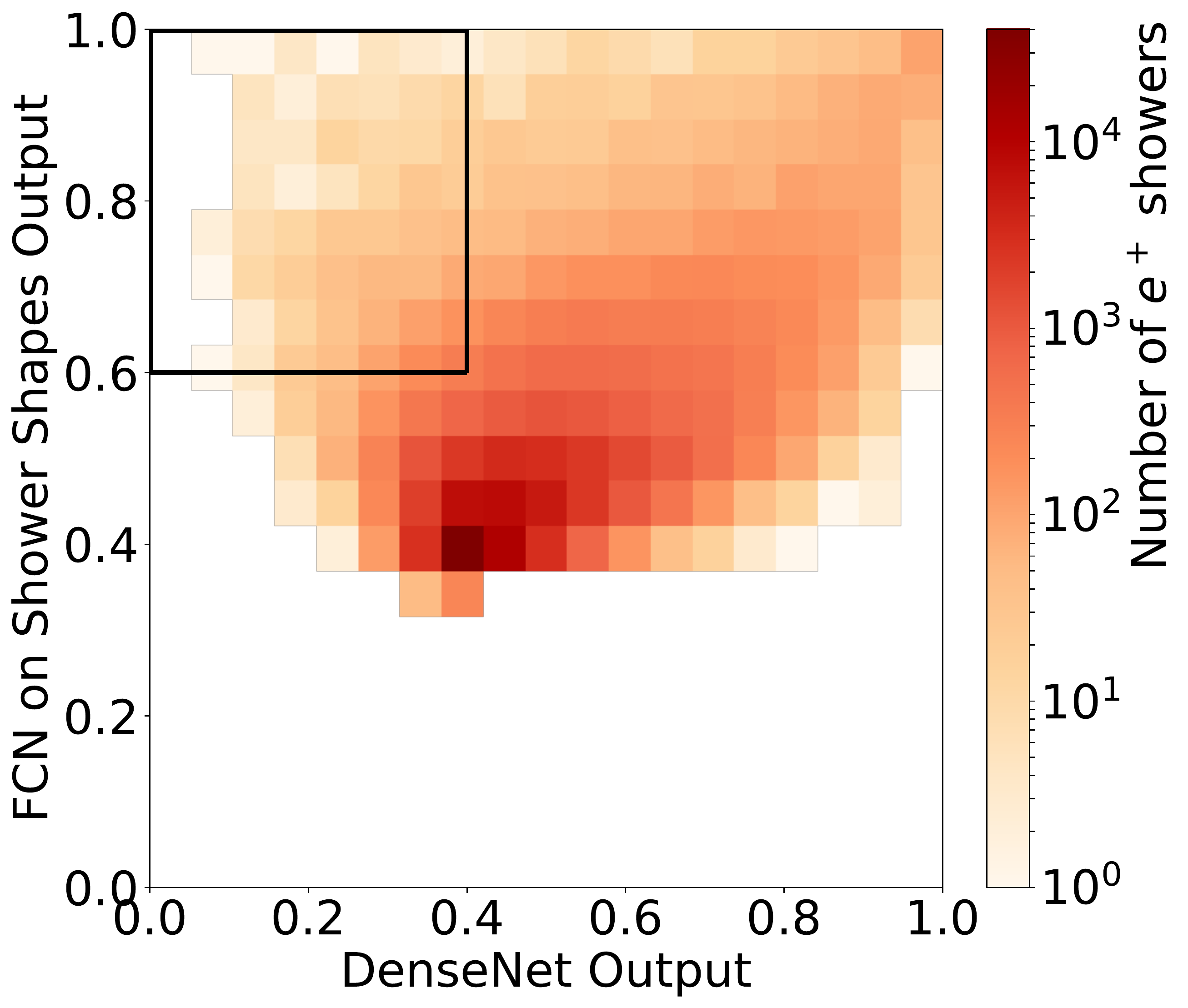}
}%
\hfill
\subfigure[2D distribution of output scores for $\gamma$ showers (target: 1). The box highlight a subset of showers that are correctly classified by the DenseNet and incorrectly classified by the shower shapes-based network.]{
\centering
        \includegraphics[width=0.45\textwidth]{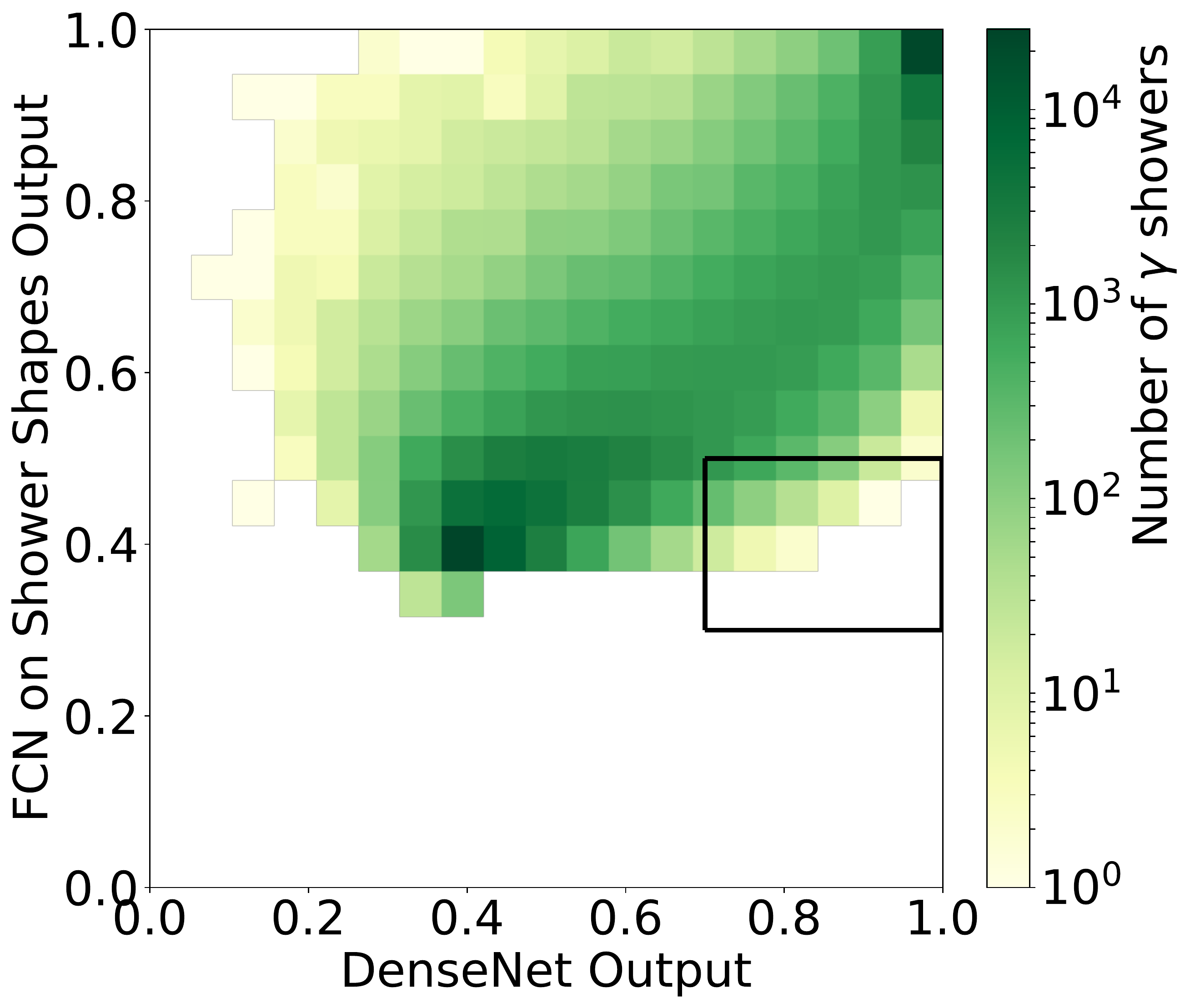}
}
\caption{Shower classification scores from the DenseNet and the FCN on shower shapes, for true $e^+$ showers on the left and true $\gamma$ showers on the right. The box in subfigure (a) highlights a region for which $\mathbb{P}_{DNet}[\gamma] < 0.4$ \& $\mathbb{P}_{FCN}[\gamma] > 0.6$, \textit{i.e.} where the DenseNet correctly assigns the positrons a low probability of being $\gamma$-originated showers, while the shower shapes method incorrectly assigns a high probability. Similarly, the box in subfigure (b) highlights a region for which $\mathbb{P}_{DNet}[\gamma] > 0.7$ \& $\mathbb{P}_{FCN}[\gamma] < 0.5$, \textit{i.e.} where the DenseNet correctly assigns the photons a high probability of being $\gamma$-originated showers, while the shower shapes method incorrectly assigns a low probability.}
\label{fig:boxes}
\end{figure}

We first plot 2-dimensional histogram of discriminants from the DenseNet~\ref{ssec:3sdnet} and the fully-connected network on shower shapes~\ref{ssec:ss} in Fig.~\ref{fig:boxes}. This allows us to identify pathological regions in which the ordinary shower shapes-based model fails to correctly classify particles, while the DenseNet succeeds at its task. 

\begin{figure}
\centering
\subfigure[Fraction of shower energy deposited in the 0$^{\mathrm{th}}$ layer]{
\centering
        \includegraphics[width=0.22\textwidth]{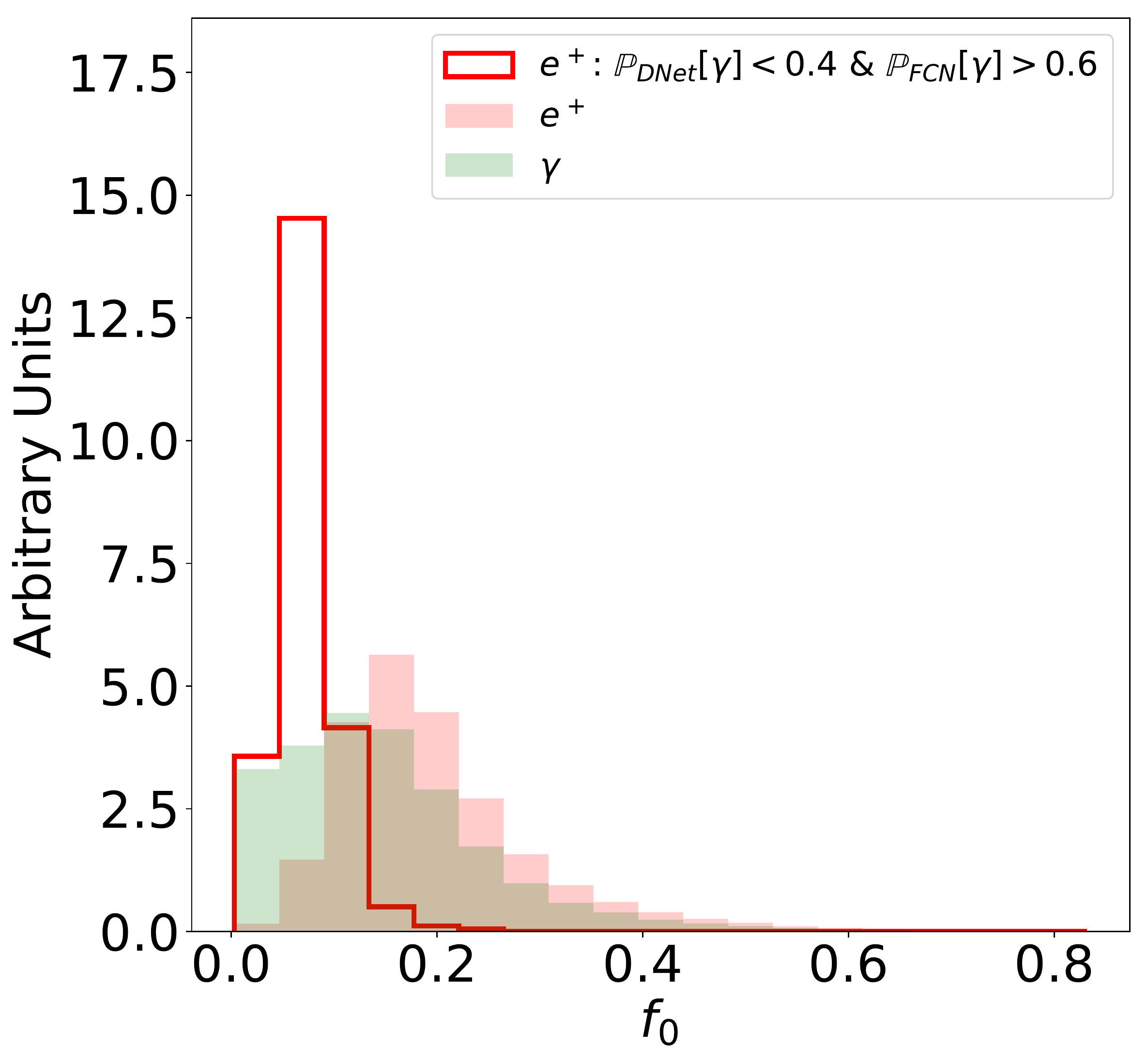}
}%
\hfill
\subfigure[Fraction of shower energy deposited in the 1$^{\mathrm{st}}$ layer]{
\centering
        \includegraphics[width=0.22\textwidth]{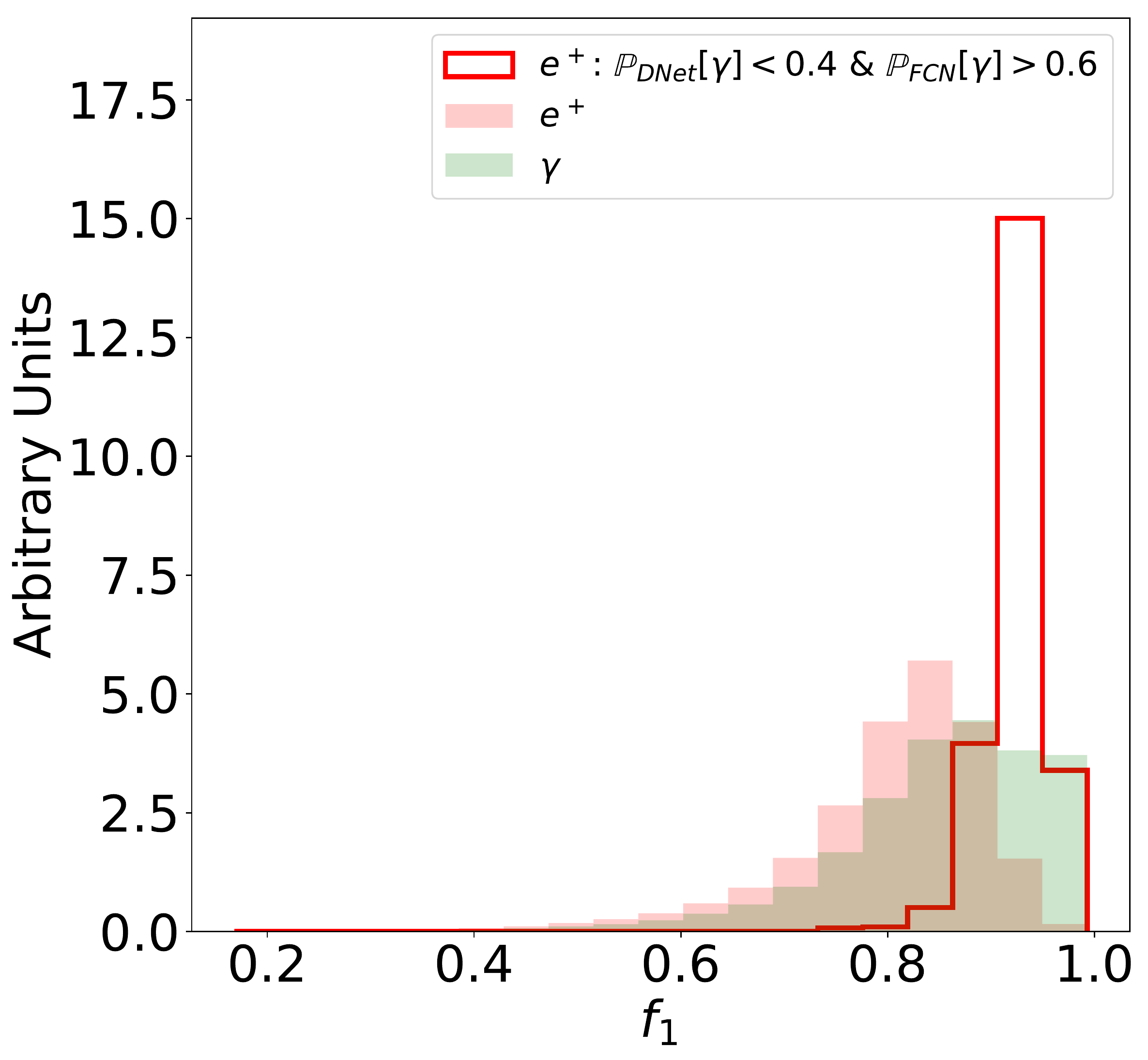}
}%
\hfill
\subfigure[Fraction of shower energy deposited in the 2$^{\mathrm{nd}}$ layer]{
\centering
        \includegraphics[width=0.22\textwidth]{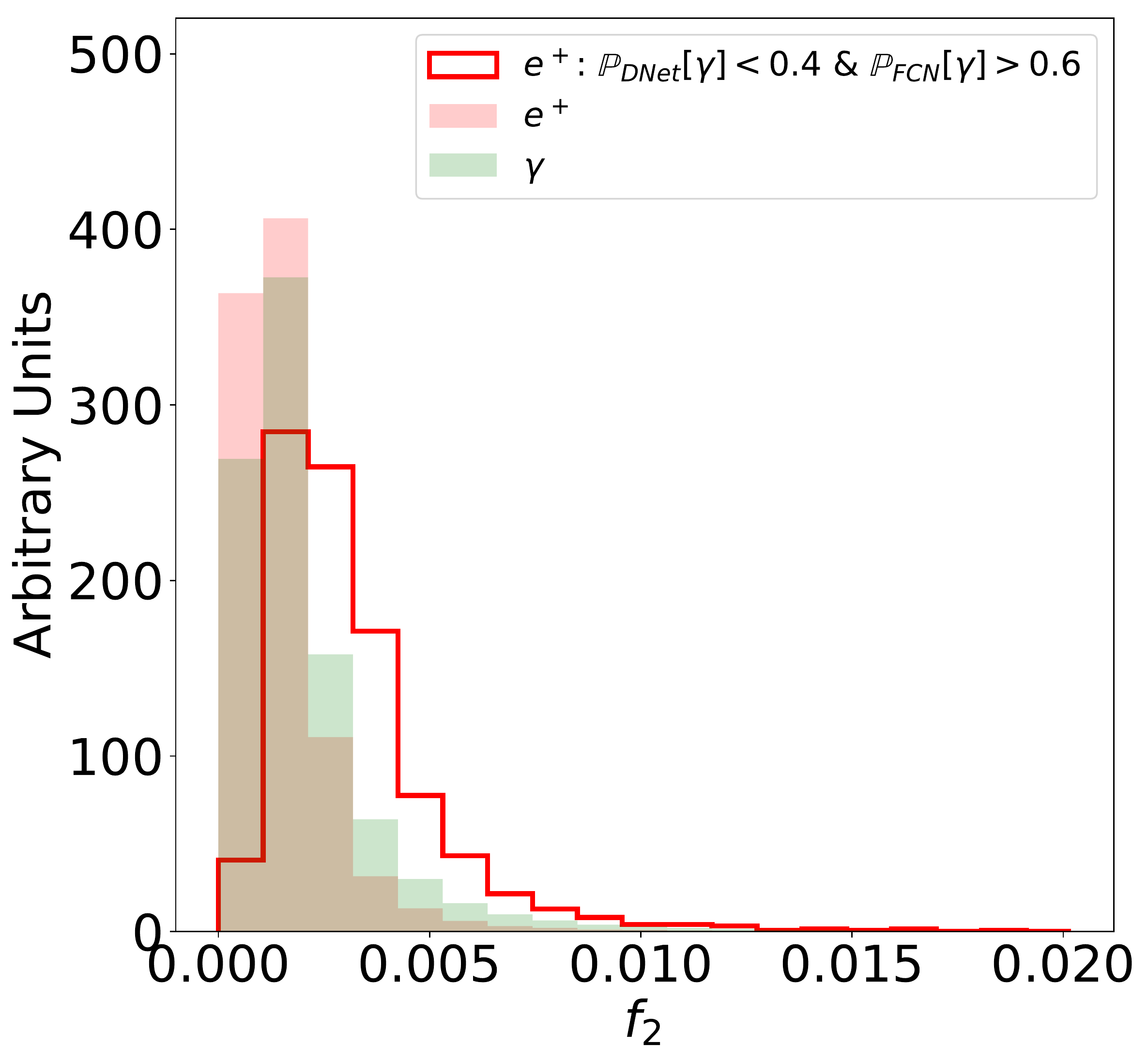}
}%
\hfill
\subfigure[Total energy]{
\centering
        \includegraphics[width=0.22\textwidth]{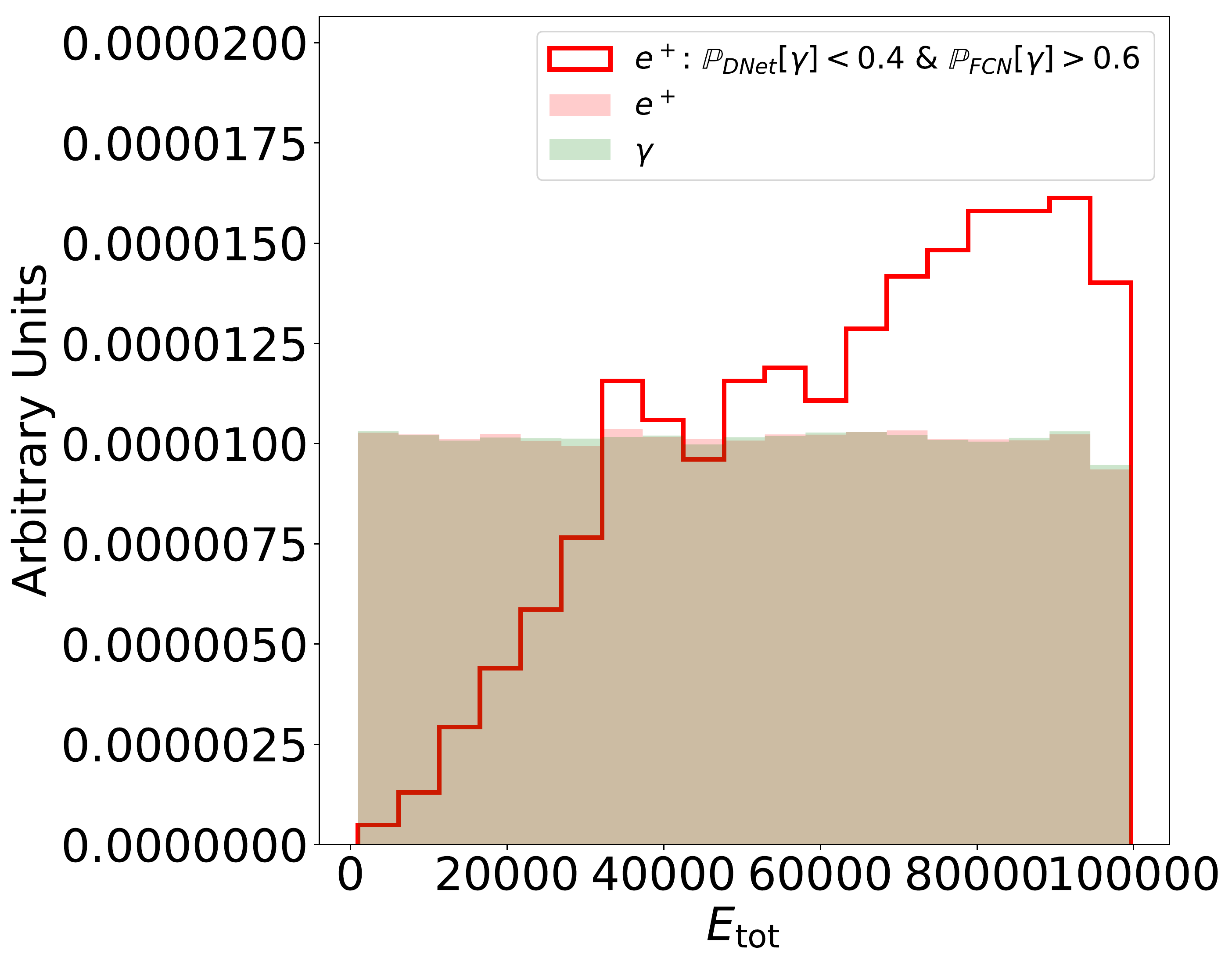}
}%
\hfill
\subfigure[Lateral width in in the 0$^{\mathrm{th}}$ layer]{
\centering
        \includegraphics[width=0.22\textwidth]{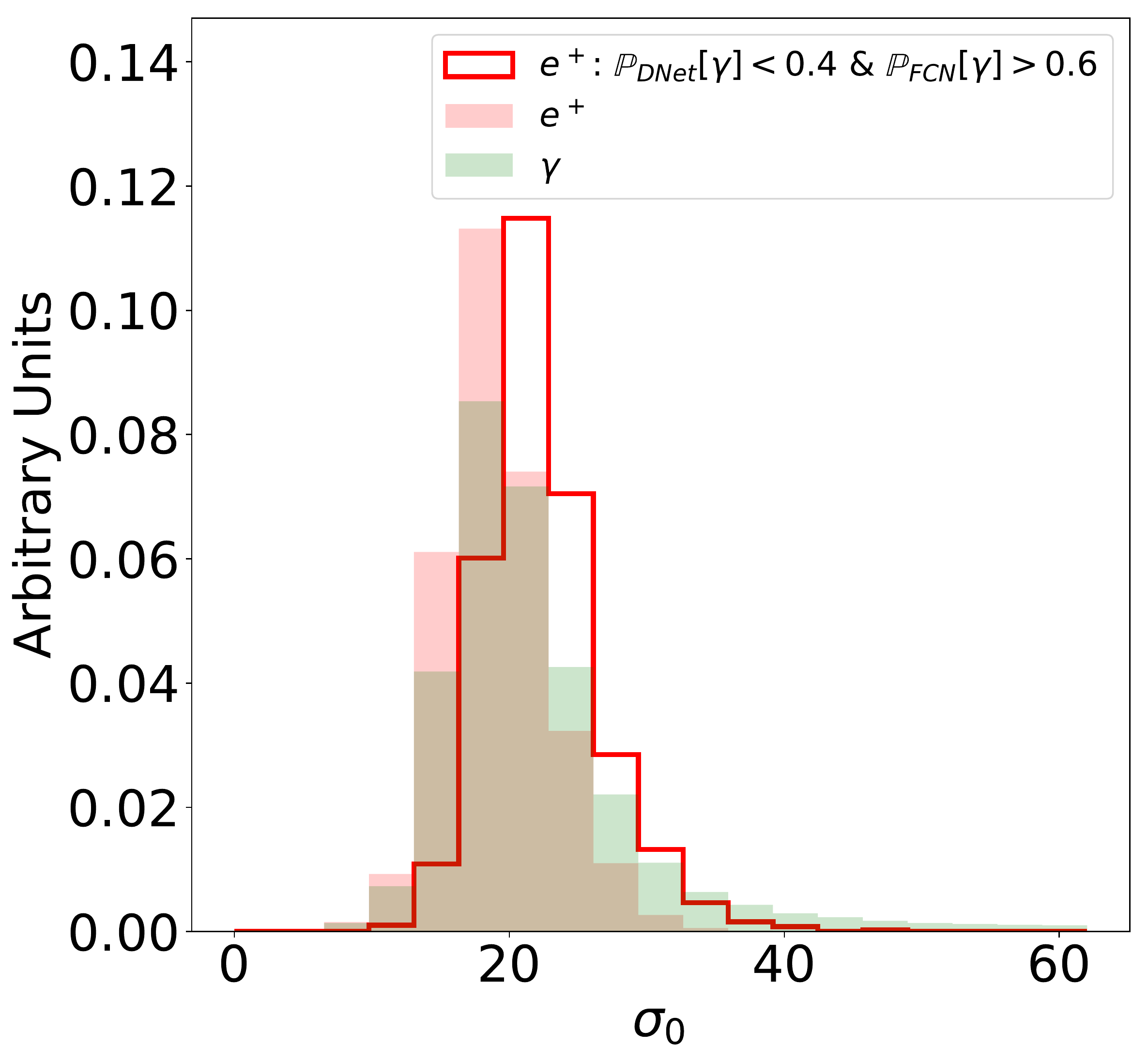}
}%
\hfill
\subfigure[Lateral width in in the 1$^{\mathrm{st}}$ layer]{
\centering
        \includegraphics[width=0.22\textwidth]{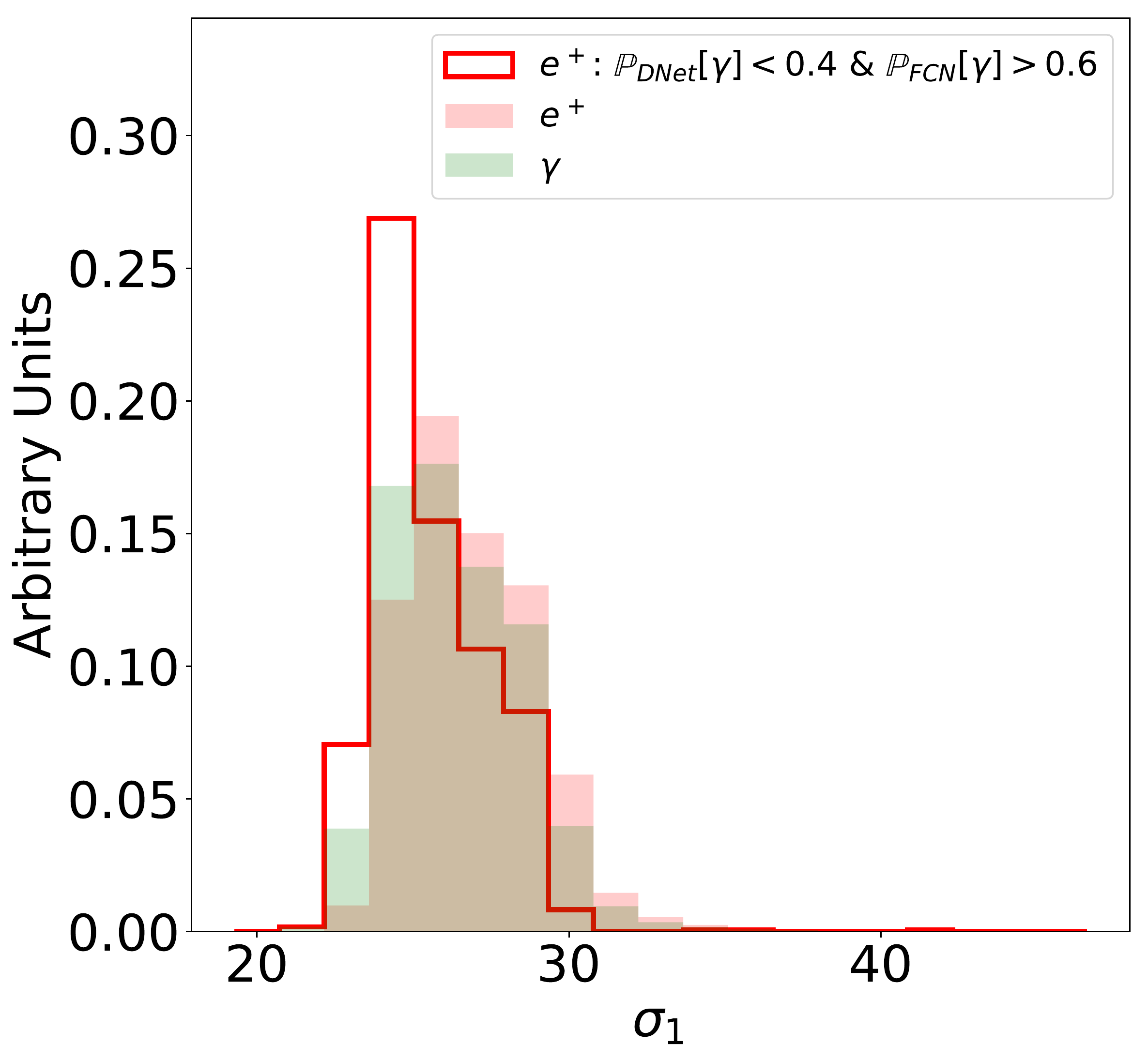}
}%
\hfill
\subfigure[Lateral depth]{
\centering
        \includegraphics[height=0.2\textwidth]{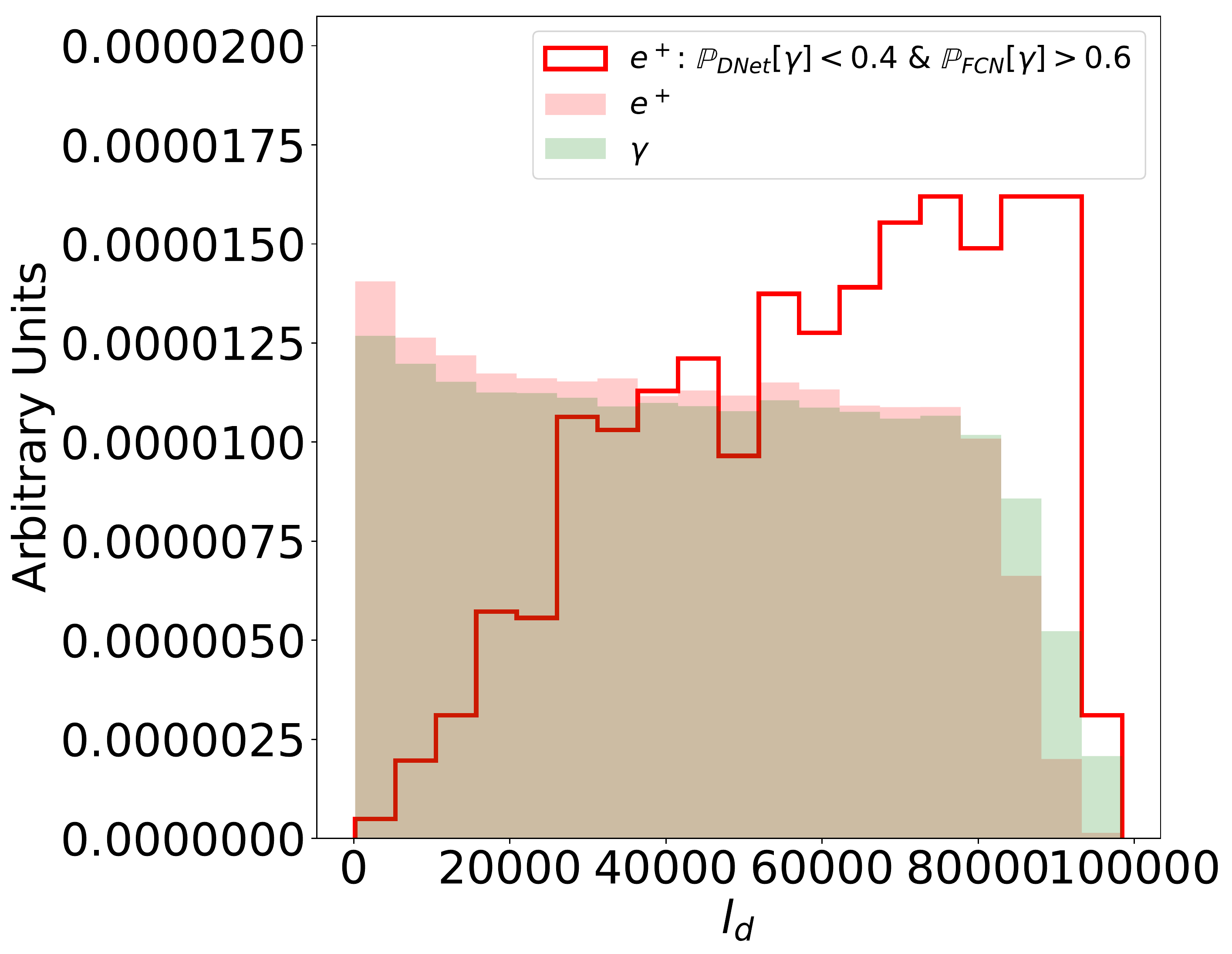}
}%
\hfill
\subfigure[Maximum depth]{
\centering
        \includegraphics[height=0.2\textwidth]{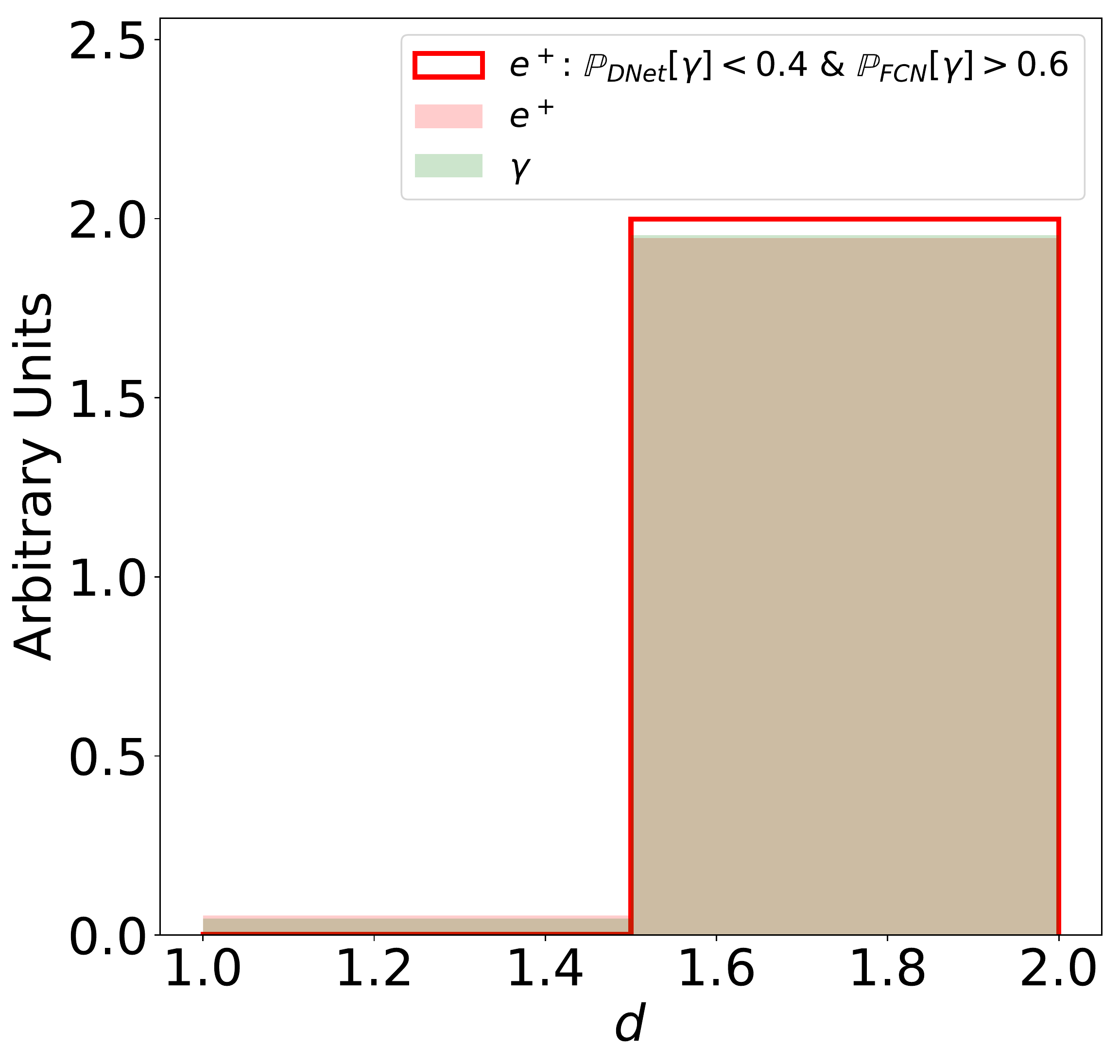}
}%
\hfill
\subfigure[Depth width]{
\centering
        \includegraphics[height=0.2\textwidth]{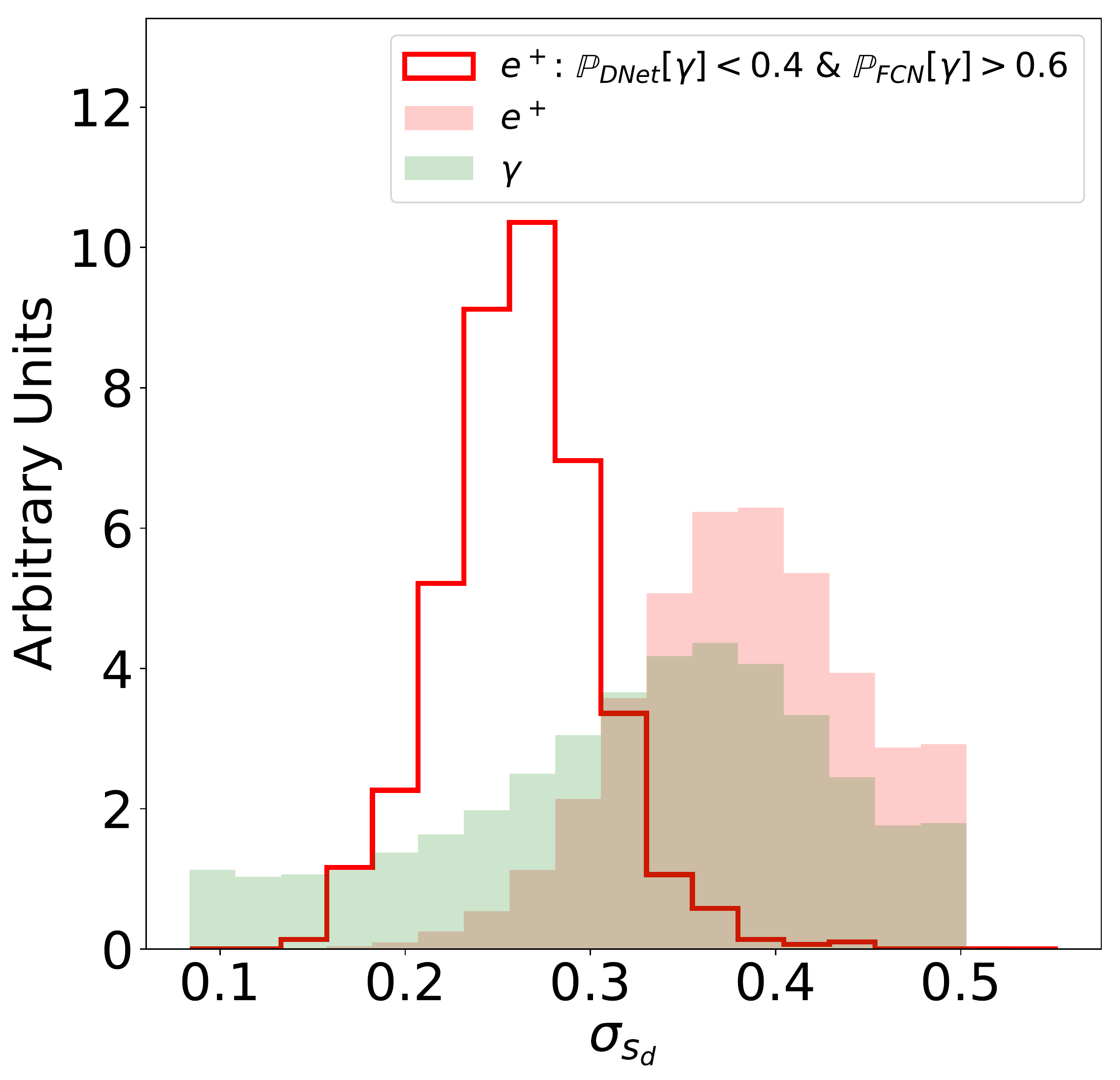}
}%
\hfill
\subfigure[Sparsity in the 0$^{\mathrm{th}}$ layer]{
\centering
        \includegraphics[height=0.2\textwidth]{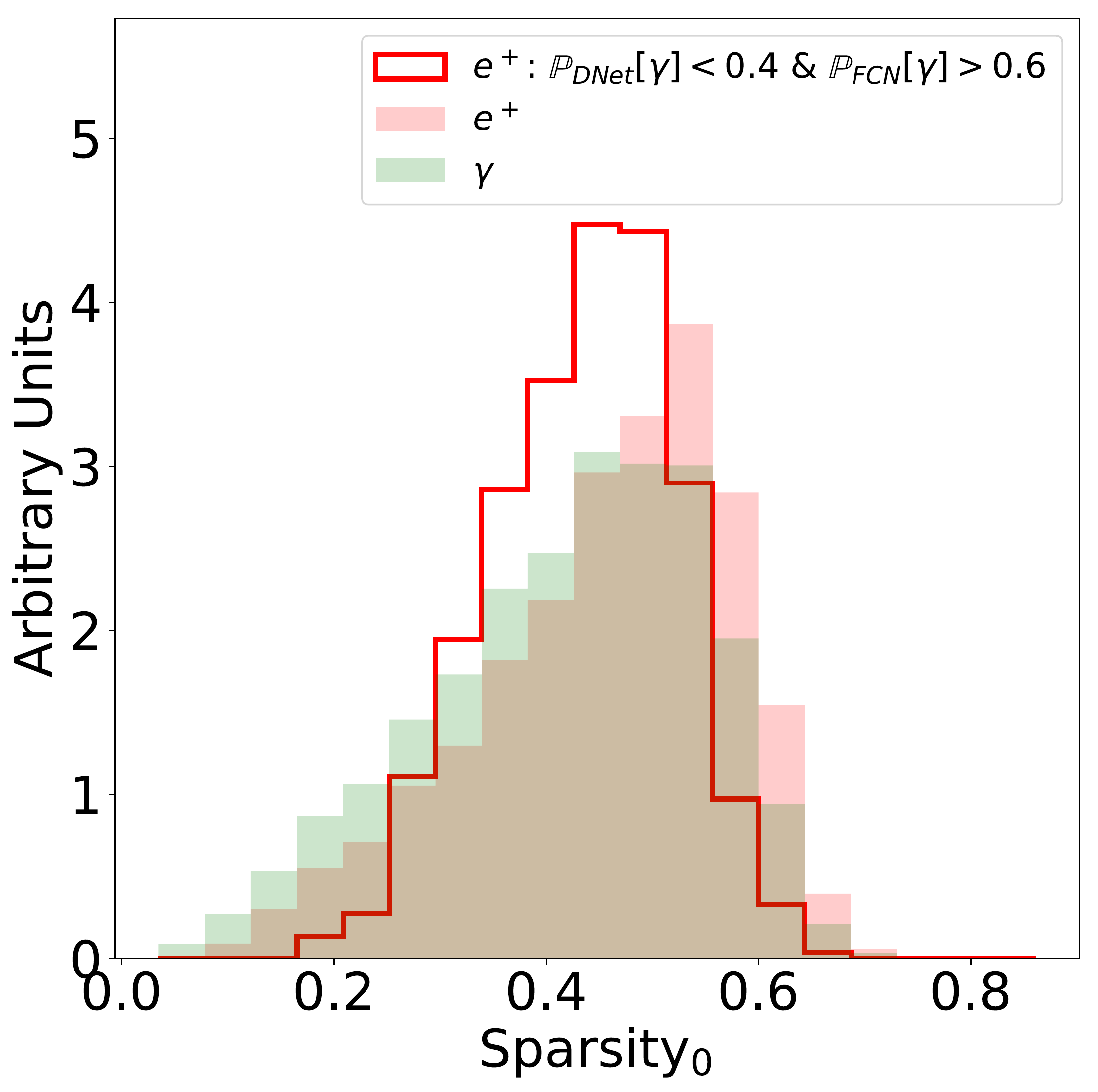}
}%
\hfill
\subfigure[Sparsity in the 1$^{\mathrm{st}}$ layer]{
\centering
        \includegraphics[height=0.2\textwidth]{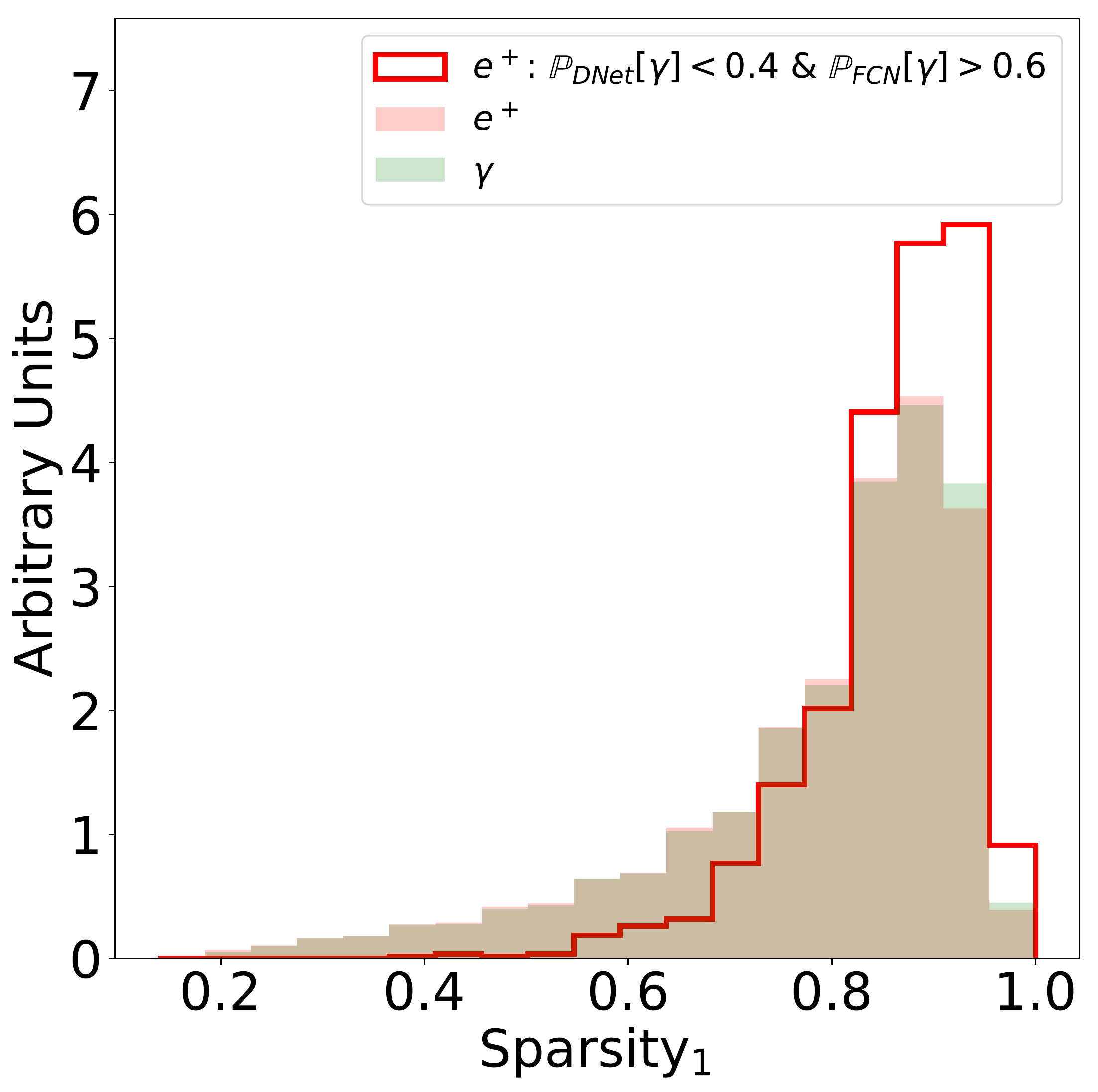}
}%
\hfill
\subfigure[Sparsity in the 2$^{\mathrm{nd}}$ layer]{
\centering
        \includegraphics[height=0.2\textwidth]{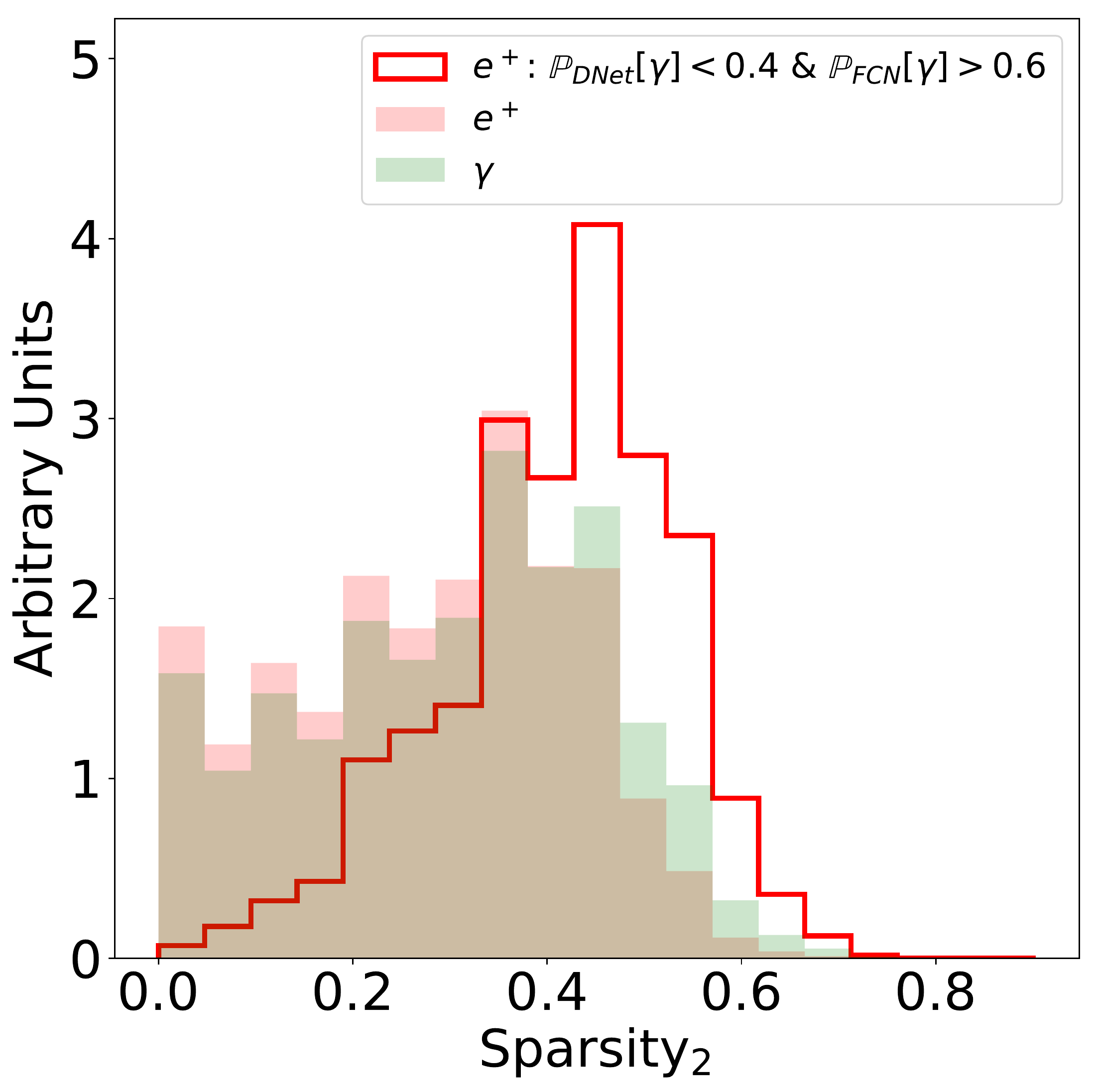}
}%
\caption{Collection of shower shape distributions in which positrons with $\mathbb{P}_{DNet}[\gamma] < 0.4$ \& $\mathbb{P}_{FCN}[\gamma] > 0.6$ (red contour) display many of the properties more typical of photons (shaded in green) than of positrons (shaded in red), which therefore causes the shower shape-based tagger to misclassify them. To compare the shape of distributions, all histograms are normalized to unit area. The positrons in the disagreement region are a subset of all positrons.}
\label{fig:ssfail_e+}
\end{figure}

\begin{figure}
\centering
\subfigure[Fraction of shower energy deposited in the 0$^{\mathrm{th}}$ layer]{
\centering
        \includegraphics[width=0.22\textwidth]{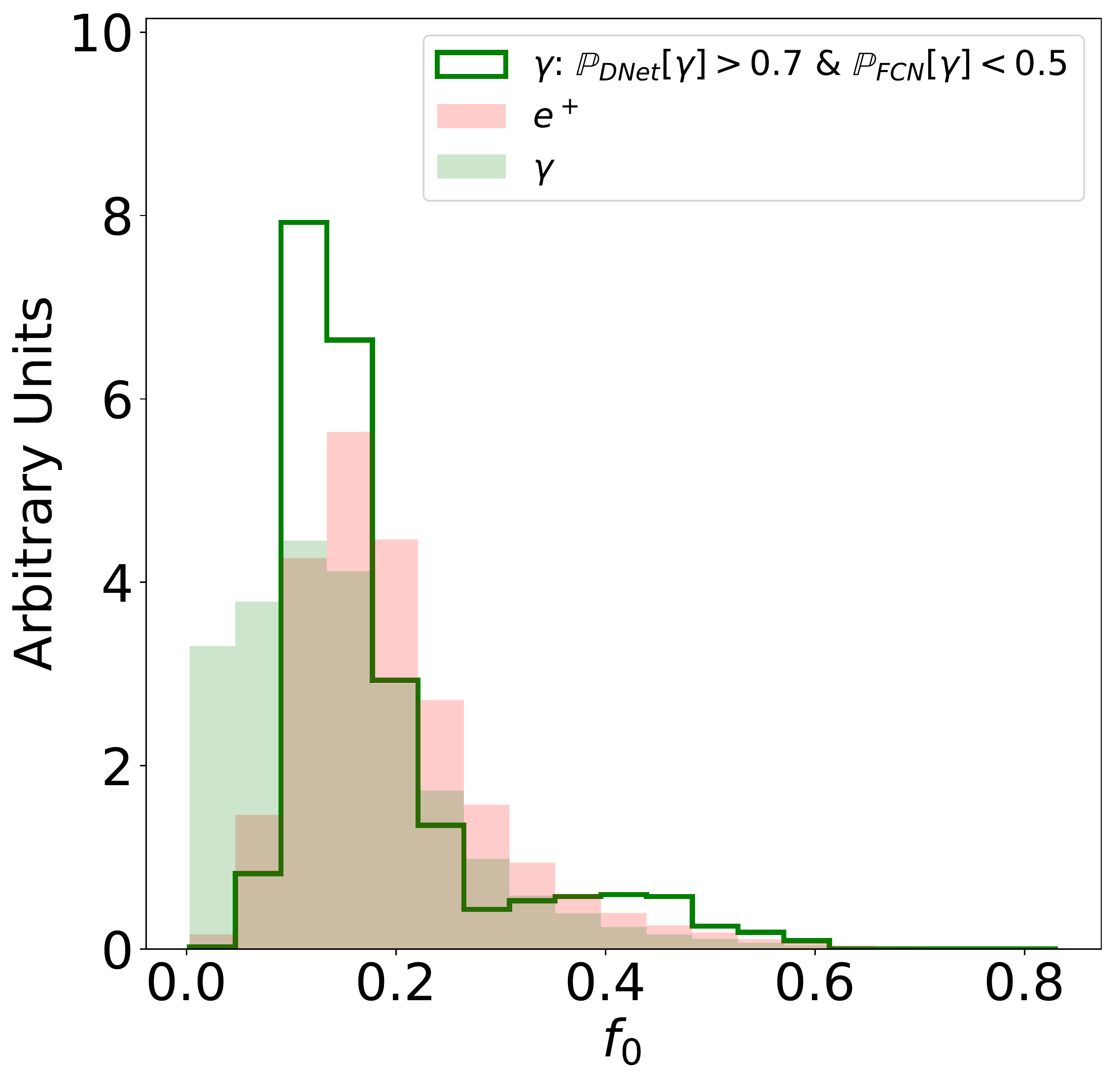}
}%
\hfill
\subfigure[Fraction of shower energy deposited in the 1$^{\mathrm{st}}$ layer]{
\centering
        \includegraphics[width=0.22\textwidth]{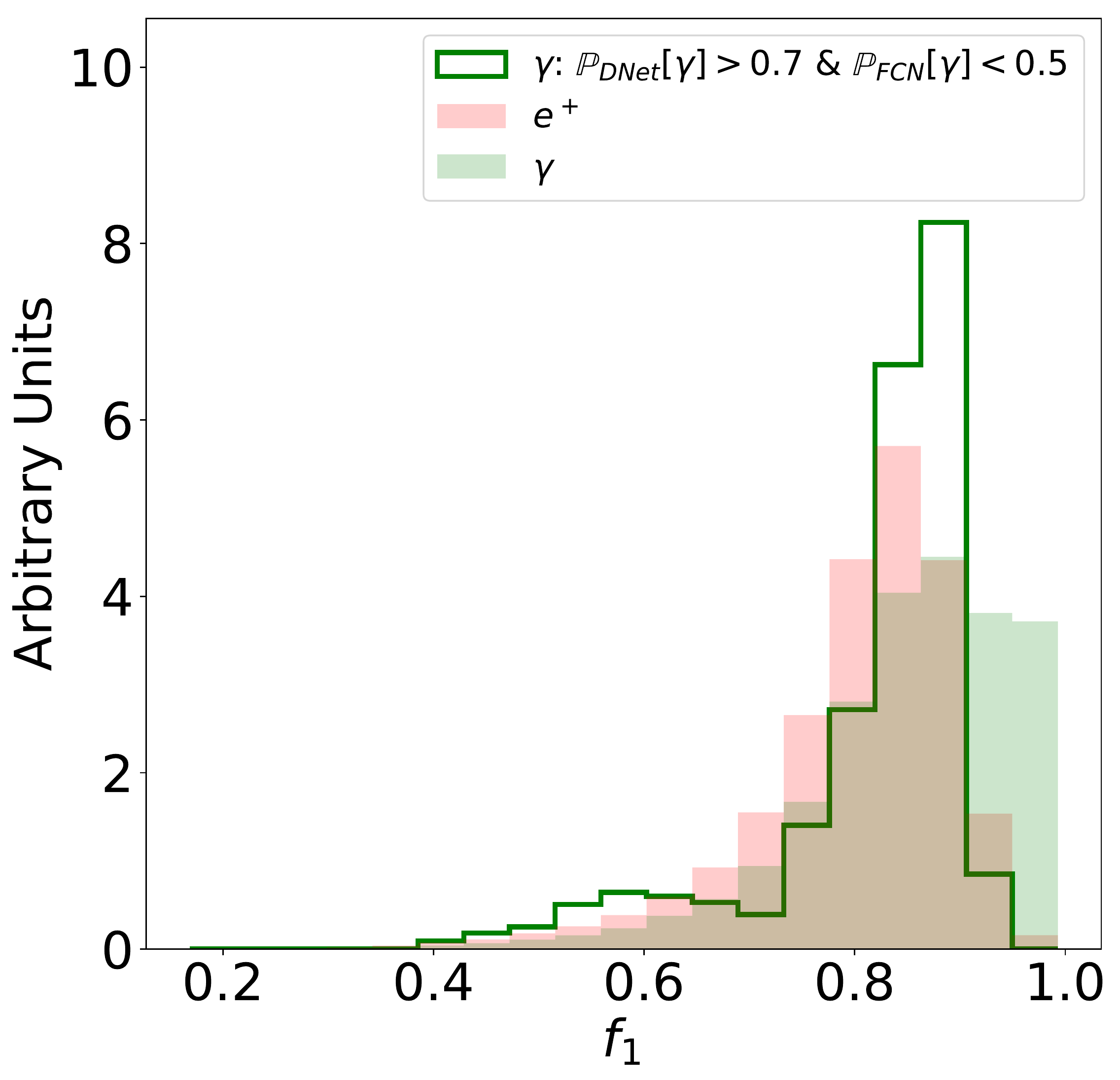}
}%
\hfill
\subfigure[Fraction of shower energy deposited in the 2$^{\mathrm{nd}}$ layer]{
\centering
        \includegraphics[width=0.22\textwidth]{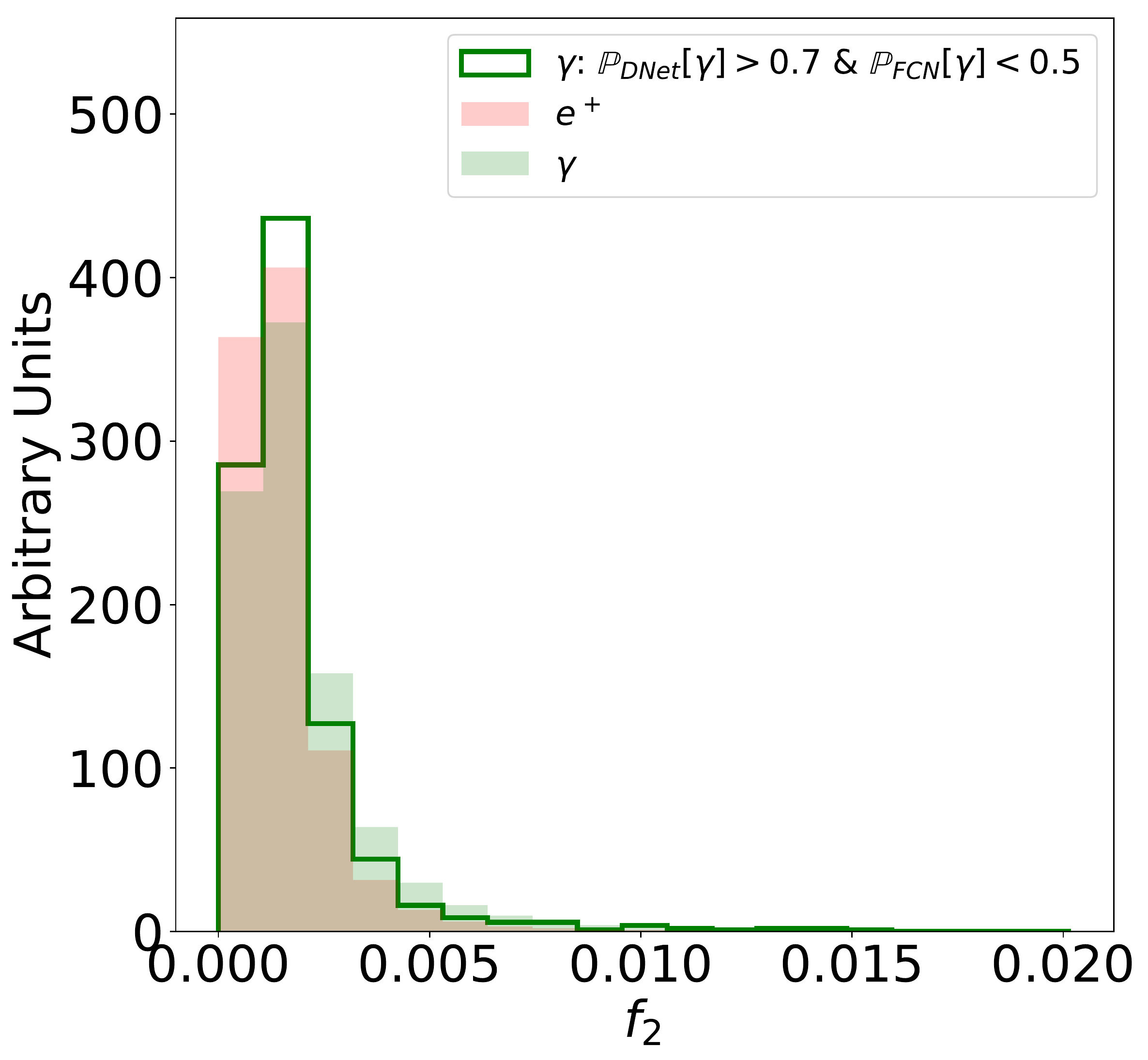}
}%
\hfill
\subfigure[Total energy]{
\centering
        \includegraphics[width=0.22\textwidth]{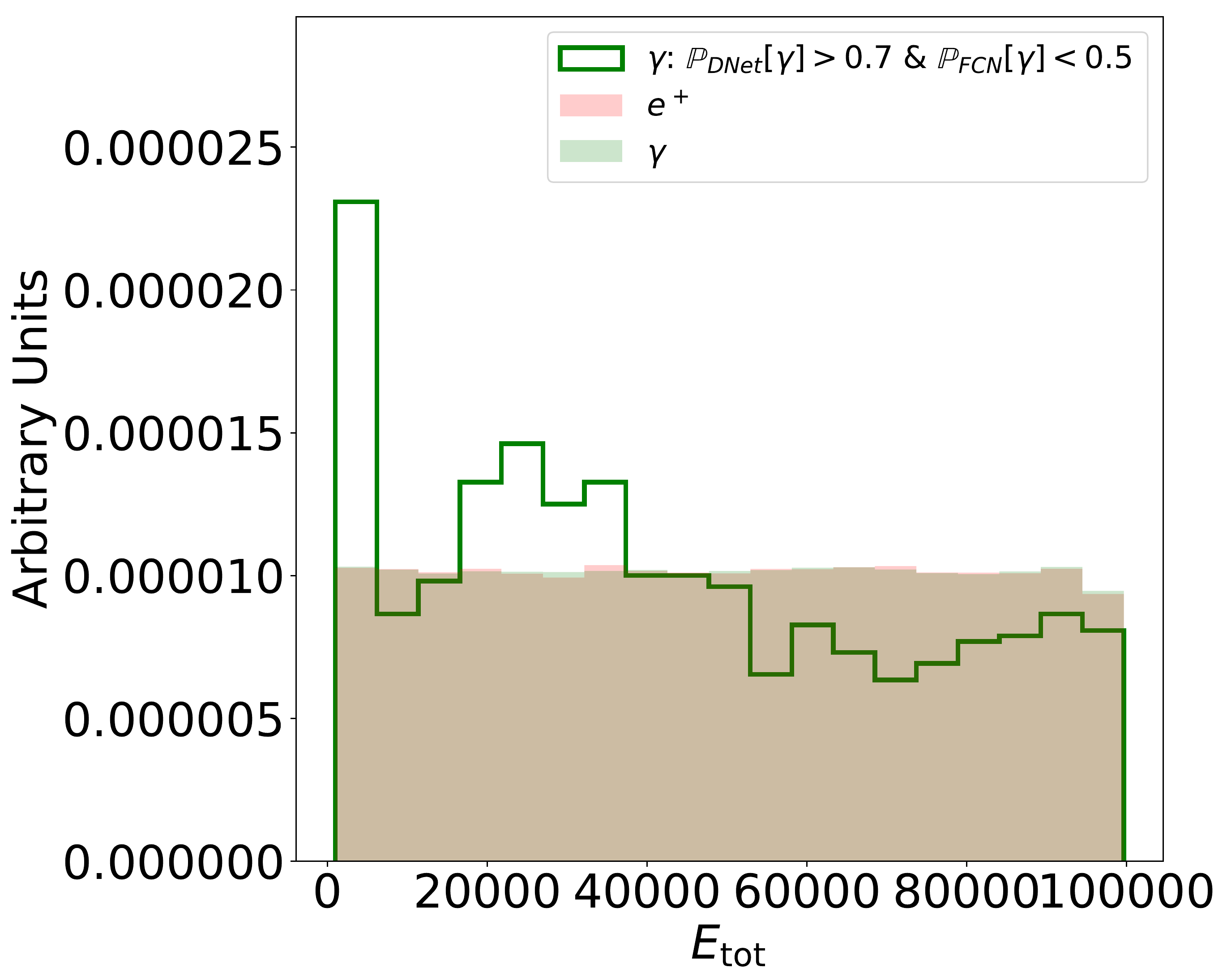}
}%
\hfill
\subfigure[Lateral width in in the 0$^{\mathrm{th}}$ layer]{
\centering
        \includegraphics[width=0.22\textwidth]{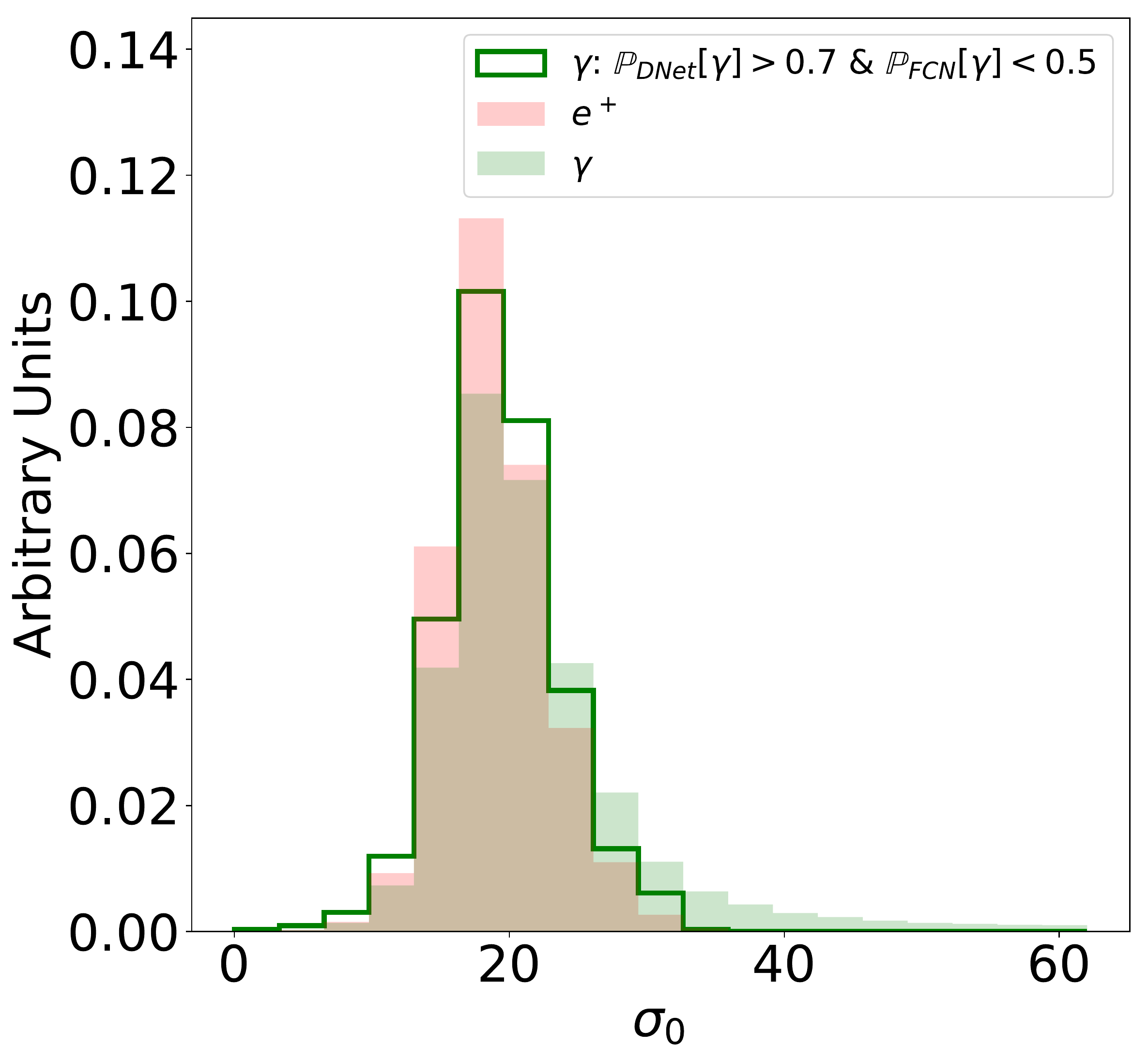}
}%
\hfill
\subfigure[Lateral depth]{
\centering
        \includegraphics[height=0.2\textwidth]{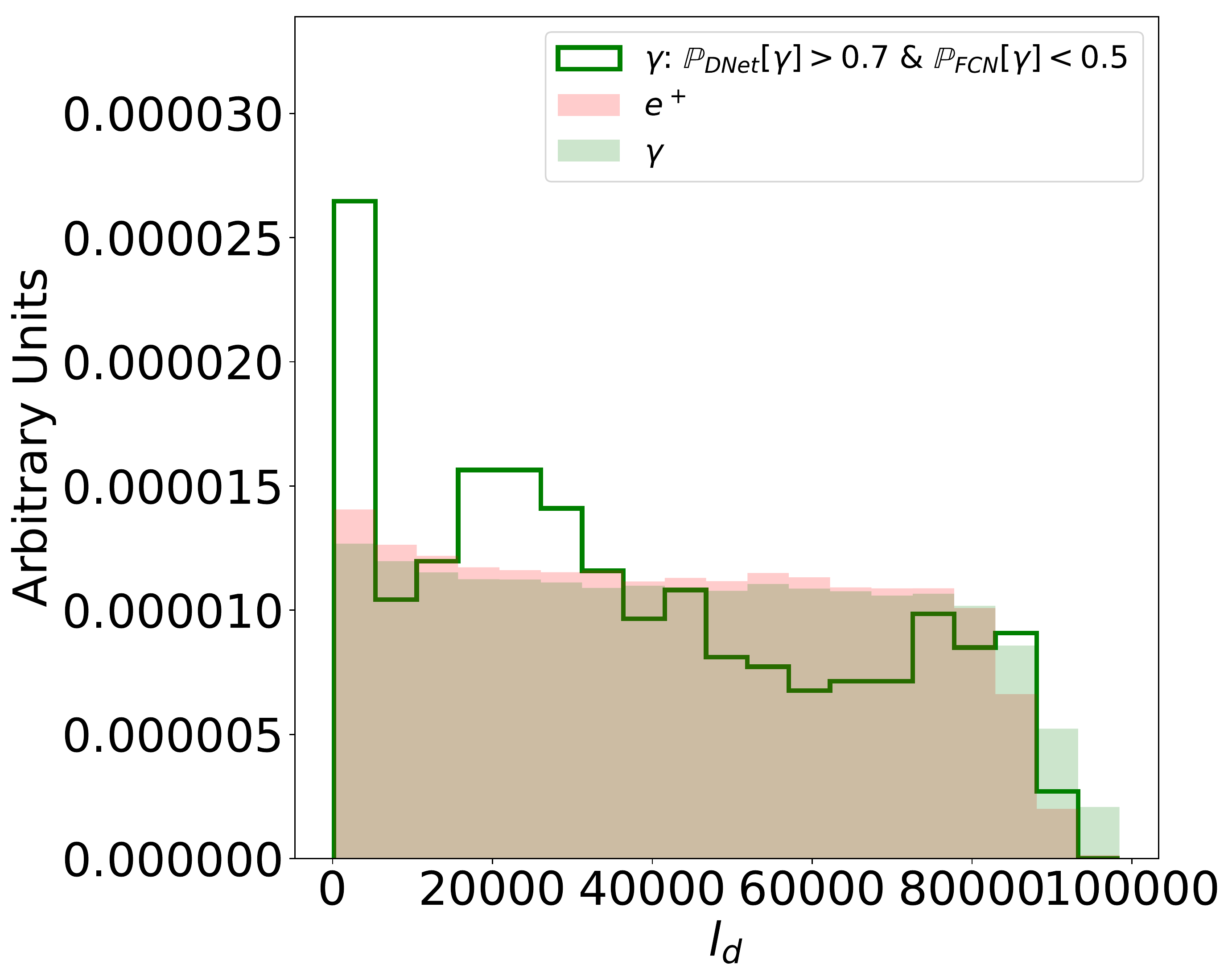}
}%
\hfill
\subfigure[Maximum depth]{
\centering
        \includegraphics[height=0.2\textwidth]{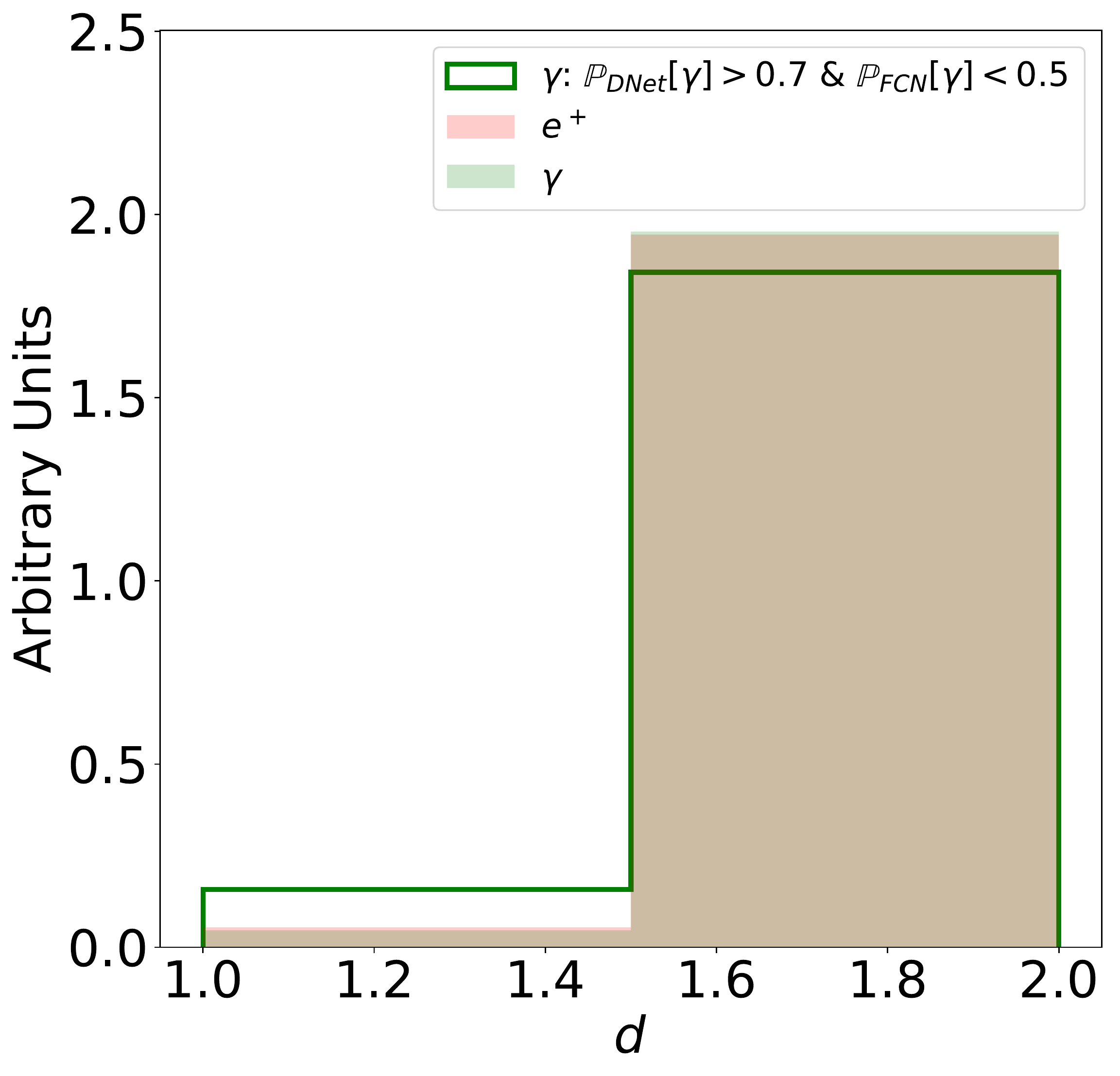}
}%
\hfill
\subfigure[Depth width]{
\centering
        \includegraphics[height=0.2\textwidth]{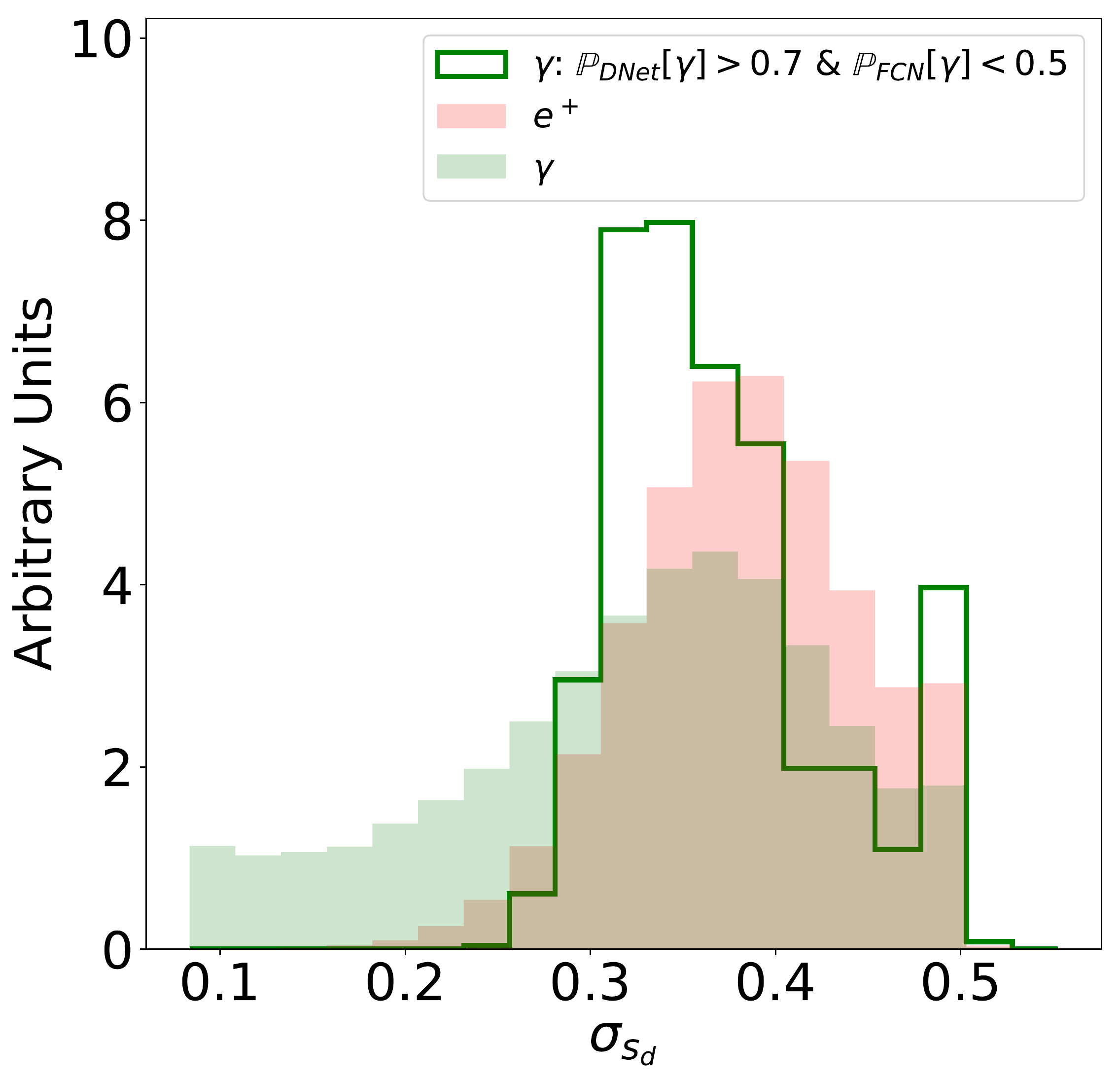}
}%

\caption{Collection of shower shape distributions in which photons with $\mathbb{P}_{DNet}[\gamma] > 0.7$ \& $\mathbb{P}_{FCN}[\gamma] < 0.5$ (green contour) display many of the properties more typical of positrons (shaded in red) than of photons (shaded in green), which therefore causes the shower shape-based tagger to misclassify them. To compare the shape of distributions, all histograms are normalized to unit area. The photons in the disagreement region are a subset of all photons.}
\label{fig:ssfail_gamma}
\end{figure}

We can further probe these regions. The shaded histograms in Figs.~\ref{fig:ssfail_e+} and ~\ref{fig:ssfail_gamma} show the distributions of shower shapes for positrons, in red, and photons, in green. In the subsets of showers that are misclassified by the baseline tagger but correctly classified by the DenseNet, the histograms are often shifted towards the mode of the distribution of the opposite class, thus explaining why the shower shape-based classifier under-performs. This provides enough evidence to conclude that the DenseNet is learning information beyond what is explained by the shower shapes, and it is correctly classifying these subsets of showers \textit{despite} their apparent similarity to the opposite class in the shower-shape basis. Further studies aimed at model interpretability would be necessary to investigate what extra knowledge the DenseNet is relying on, and how this information can be used to augment the shower shapes.

\section{Regression on Incoming Angles and Positions}
\label{sec:regression}
A regression task is performed on the photon shower data set to infer the coordinates of the particle's point of incidence with the first layer of the calorimeter ($x_0$, $y_0$), as well as its angular direction expressed in terms of the collider detector coordinates ($\phi_C$, $\theta_C$).

This task serves as a simple demonstration that incoming particle position and direction are easily recovered from our suggested image-based multi-layer calorimeter representation. Given the axis of a reconstructed shower object -- possibly from the outcome of a clustering algorithm -- these quantities could be geometrically deduced directly from the pixel intensities in different layers, within the uncertainties associated with calorimeter resolution. We bypass the object reconstruction step and instead regress on the angle and coordinates of incidence starting from the raw pixel activations. 

For fast turnaround, a simple fully-connected neural network on the unraveled pixel intensities is used in this proof of concept, though we expect convolutional architectures to be able to outperform our baseline. Nonetheless, future work can make use of this benchmark to test new methodologies.

\subsection{Model and Training Procedure}
The following neural network structure is adopted: four fully connected - LeakyReLU~\cite{leaky} - dropout~\cite{dropout} blocks, with hidden representations of size 512, 1024, 1024, and 128 respectively, plus a final four-dimensional output with linear activation. The network has a total of almost 2 million trainable parameters, and the batch size is chosen to be 128.

As in the previous task, the net is built using \textsc{Keras} with \textsc{TensorFlow} as a backend, trained on a Intel$^\circledR$ Xeon$^\circledR$ CPU E5-1620 v3 @ 3.50GHz with the Adam optimizer to minimize the mean absolute error between the true and predicted values. 200,000 photon showers are set aside for final testing, while 300,000 are used for the training procedure (30\% of which only serve for validation). The input pixel intensities, as well as the target output quantities $x_0$, $y_0$, $\phi_{C}$, $\theta_{C}$ are individually scaled by their highest entry in the data set. 

\subsection{Experimental Results}
\begin{figure}
\centering
\subfigure[Azimuthal angle]{
\centering
        \includegraphics[width=0.45\textwidth]{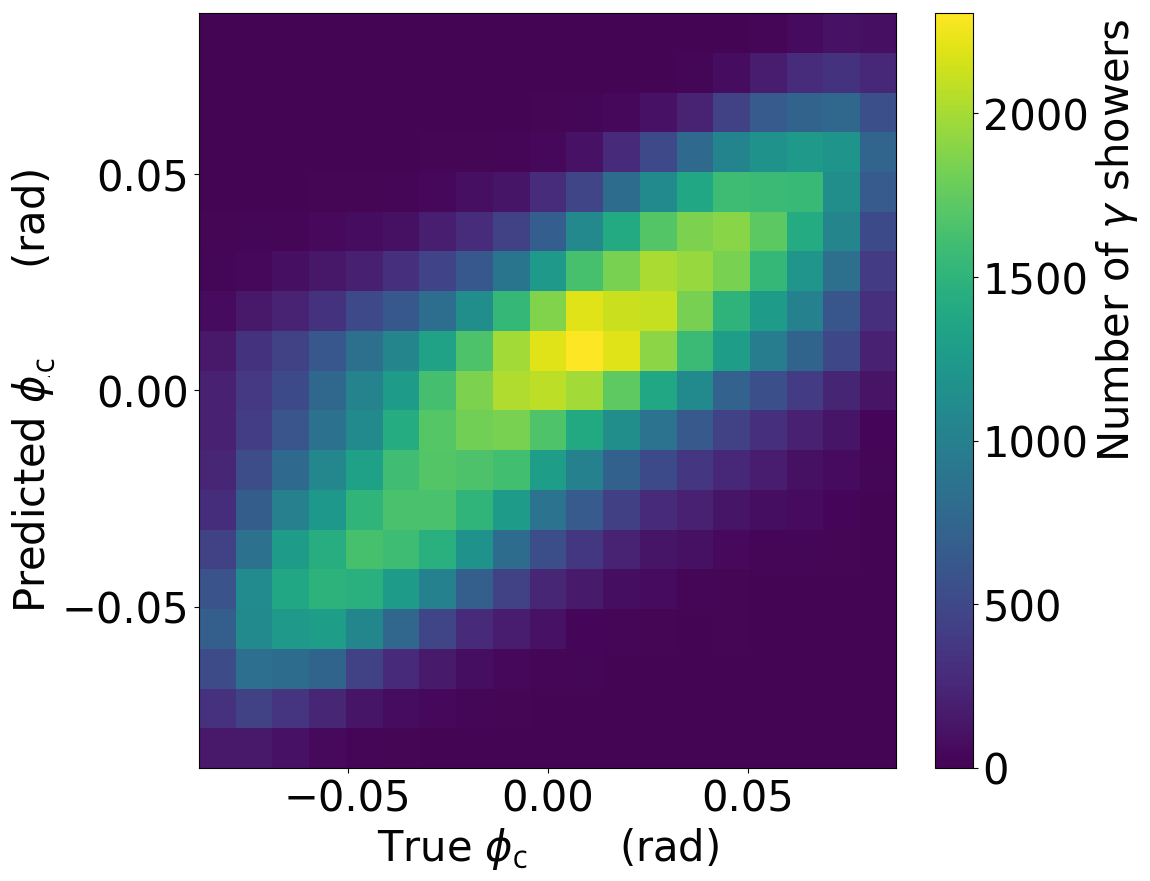}
}%
\subfigure[Polar angle]{
\centering
        \includegraphics[width=0.45\textwidth]{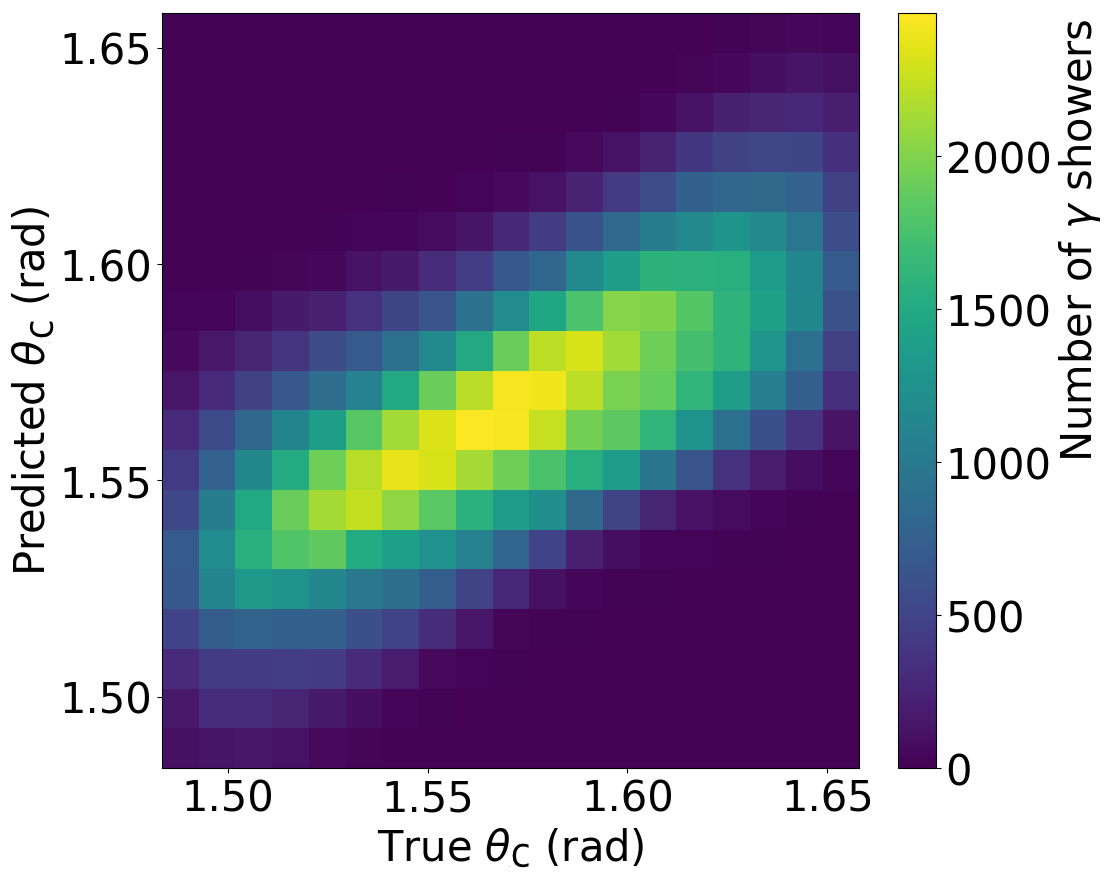}
}
\subfigure[Displacement in the coarsely segmented direction]{
\centering
        \includegraphics[width=0.45\textwidth]{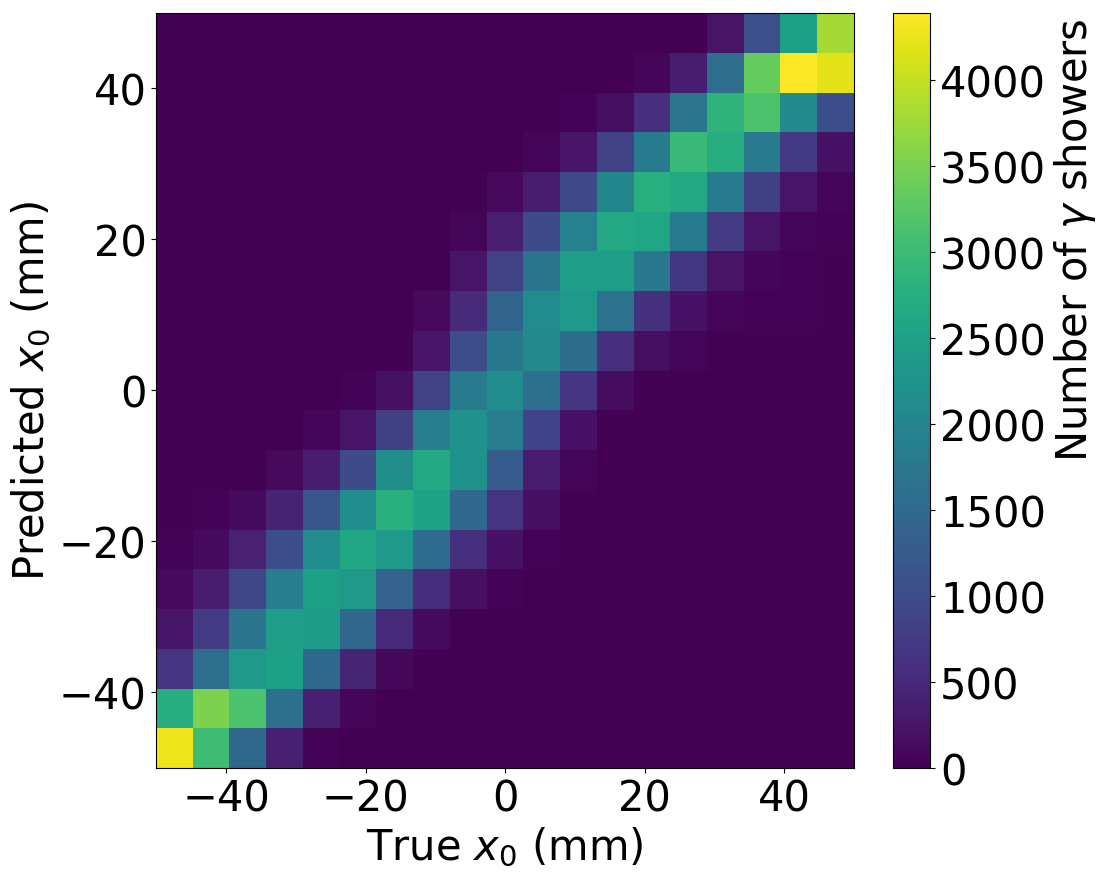}
}
\subfigure[Displacement in the finely segmented direction]{
\centering
        \includegraphics[width=0.45\textwidth]{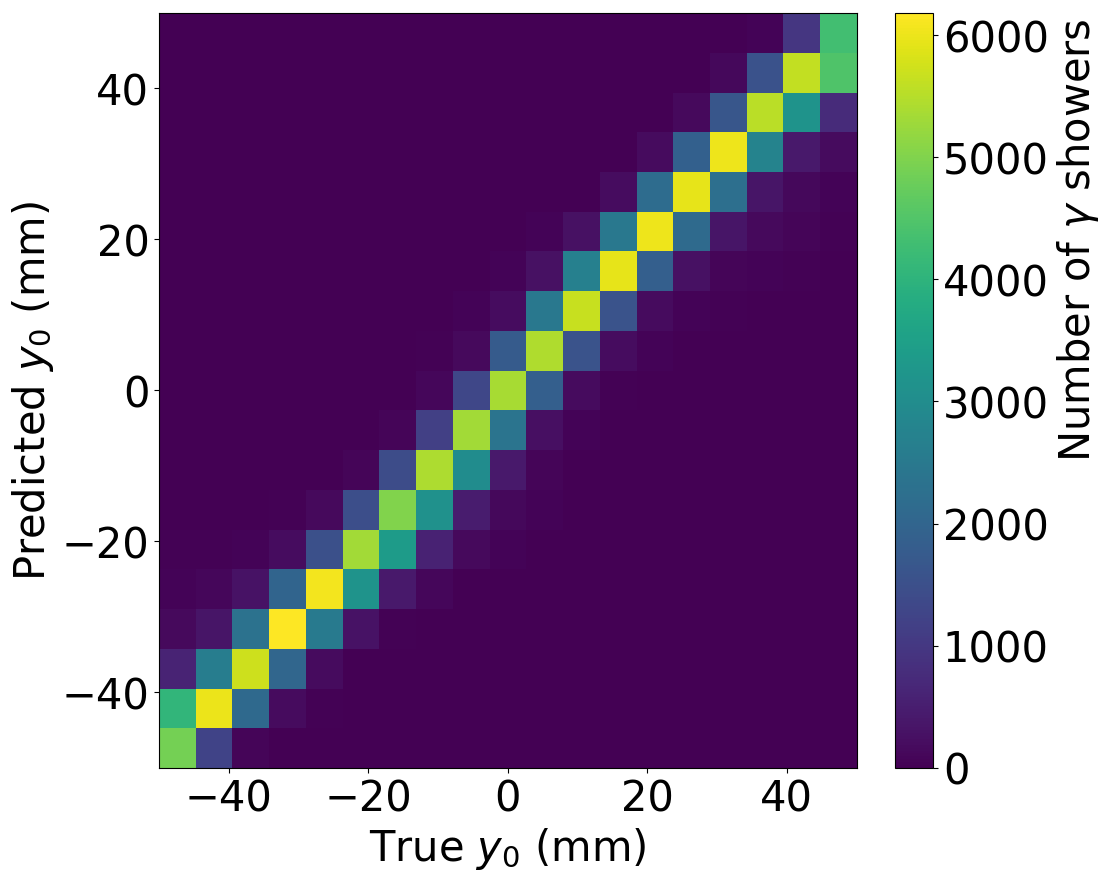}
}
\caption{Regression results on the four variables that define the incident direction and position of photons upon contact with the first layer of the electromagnetic calorimeter.}
\label{fig:regression}
\end{figure}

The mean absolute error (MAE) and root mean square error (RMSE) are reported for each regression task in Table~\ref{table:regression}, while Fig.~\ref{fig:regression} shows the unnormalized 2-dimensional histograms for true and predicted values of incident angles and positions.

It is noteworthy that the mean absolute errors of the regressions on the coordinates of the location of incidence $(x_0, y_0)$ are smaller than the cell dimensions in the first layer. 

\begin{table}[]
\centering
\caption{Regression evaluation metrics on a test set of 200,000k unseen $\gamma$ showers}
\label{table:regression}
\begin{tabular}{cll}
\toprule
Variable (units)                & MAE & RMSE \\
\midrule                               
$\phi_{C}$ (rad)                & 0.024 & 0.030 \\
$\theta_{C}$ (rad)              & 0.026 & 0.032 \\
$x_0$ (mm)                      & 6.162 & 7.876 \\
$y_0$ (mm)                      & 2.959 & 4.221 \\
\bottomrule
\end{tabular}
\end{table}

\section{Conclusion}
\label{sec:conclusion}

Building on work presented in Ref.~\cite{Paganini:2017hrr,Paganini:2017dwg, deOliveira:2017rwa}, we have studied the classification and regression performance of deep neural networks on calorimeter images. In particular, we find that the DenseNet architecture is a powerful tool for classification and is parameter efficient with respect to alternative approaches. In addition, regression tasks at the pixel level can be used to infer particle kinematics without direct physics object reconstruction, and can be integrated into the discriminator of a generative adversarial network to successfully impose constraints and condition the generative process~\cite{deOliveira:2017rwa}.

Hopefully these results will be useful when developing reconstruction algorithms for the present ATLAS detector~\cite{Aad:2008zzm}, for the future CMS detector~\cite{Collaboration:2293646}, as well as proposed future heterogeneously and longitudinally segmented calorimeter designs~\cite{calice}.

\section*{Acknowledgements}
The authors acknowledge the presence of similar ongoing work in the community and look forward to future published results. We would like to thank the participants of ACAT 2017 for interesting and useful discussions. This work was supported by the U.S.~Department of Energy, Office of Science under contracts DE-AC02-05CH11231, DE-SC0017660.

\clearpage

\bibliographystyle{report}
\bibliography{mybib.bib}

\clearpage

\appendix
\section{Dataset Production and Coordinate Definition}
\label{appA}

\begin{figure}
    \centering
    \includegraphics[width=0.5\textwidth]{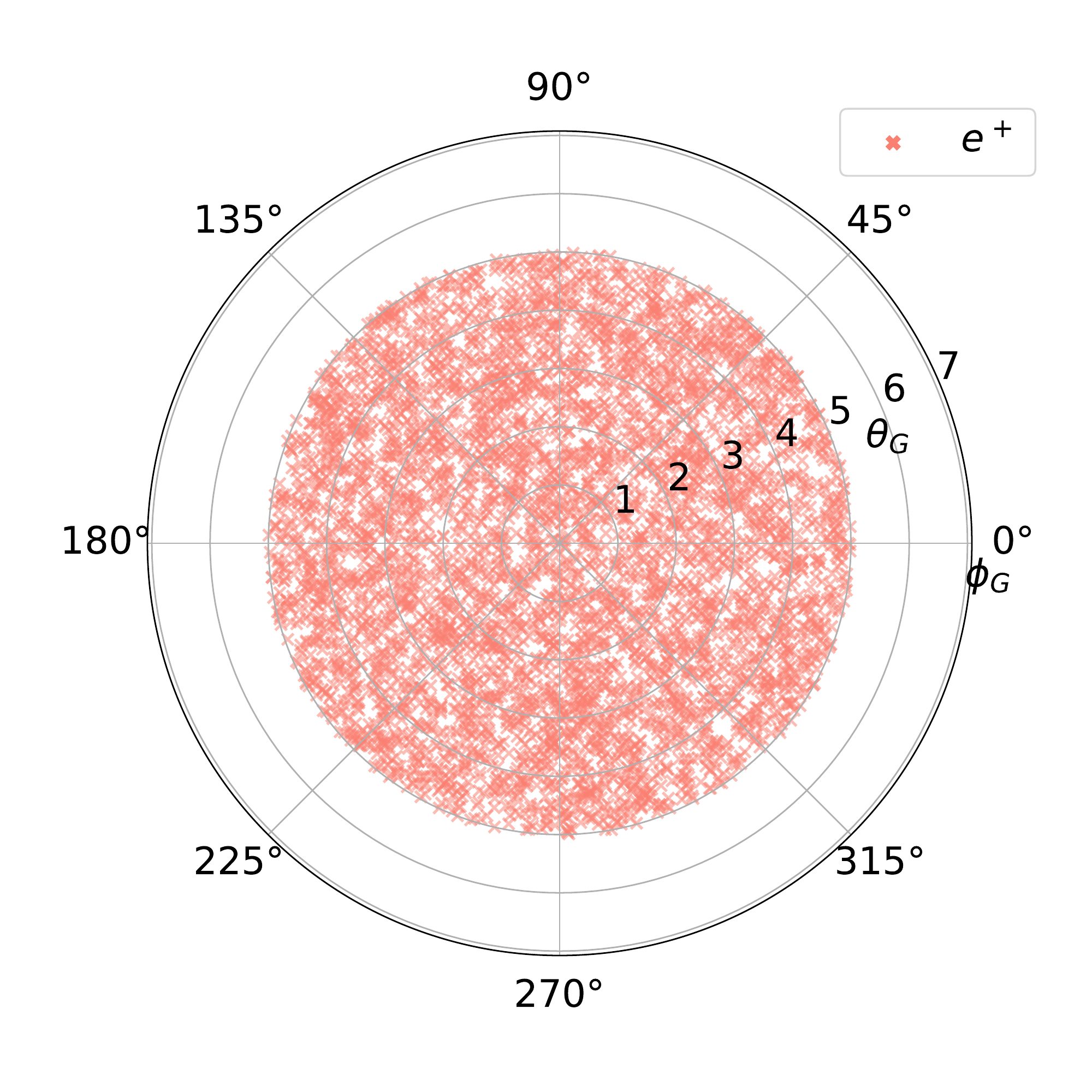}
    \caption{Isotropic distribution of 5,000 positrons from the training dataset. The concentric marks indicate the extent of the angle $\theta_G$, while the distribution is uniform across $360^\circ$ in $\phi_G$. The $\hat{z}_{G}$-axis is into the page. }
    \label{fig:target}
\end{figure}

\begin{figure}
\centering
\subfigure[Azimuthal angle]{
\centering
        \includegraphics[width=0.45\textwidth]{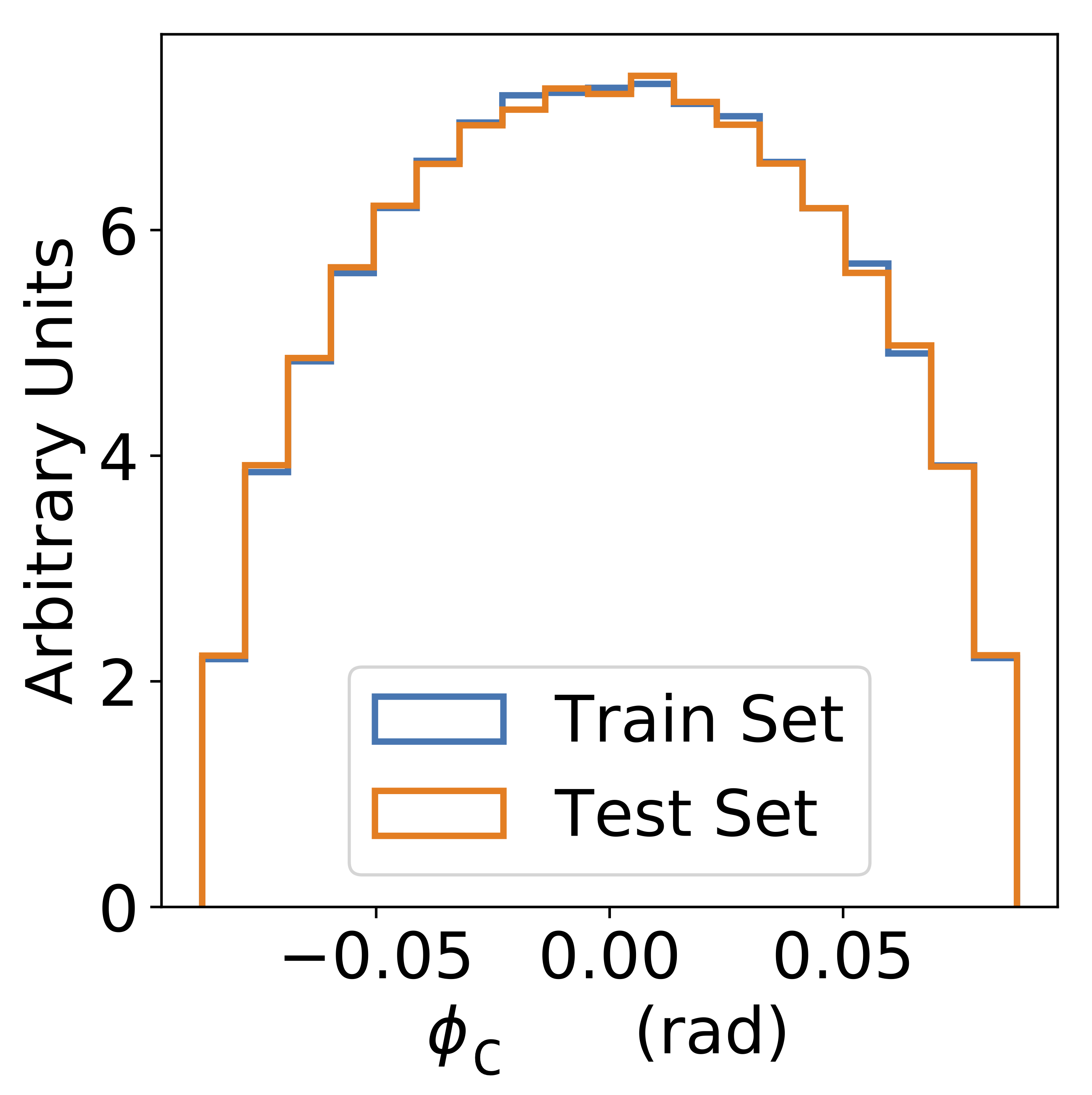}
}%
\subfigure[Polar angle]{
\centering
        \includegraphics[width=0.45\textwidth]{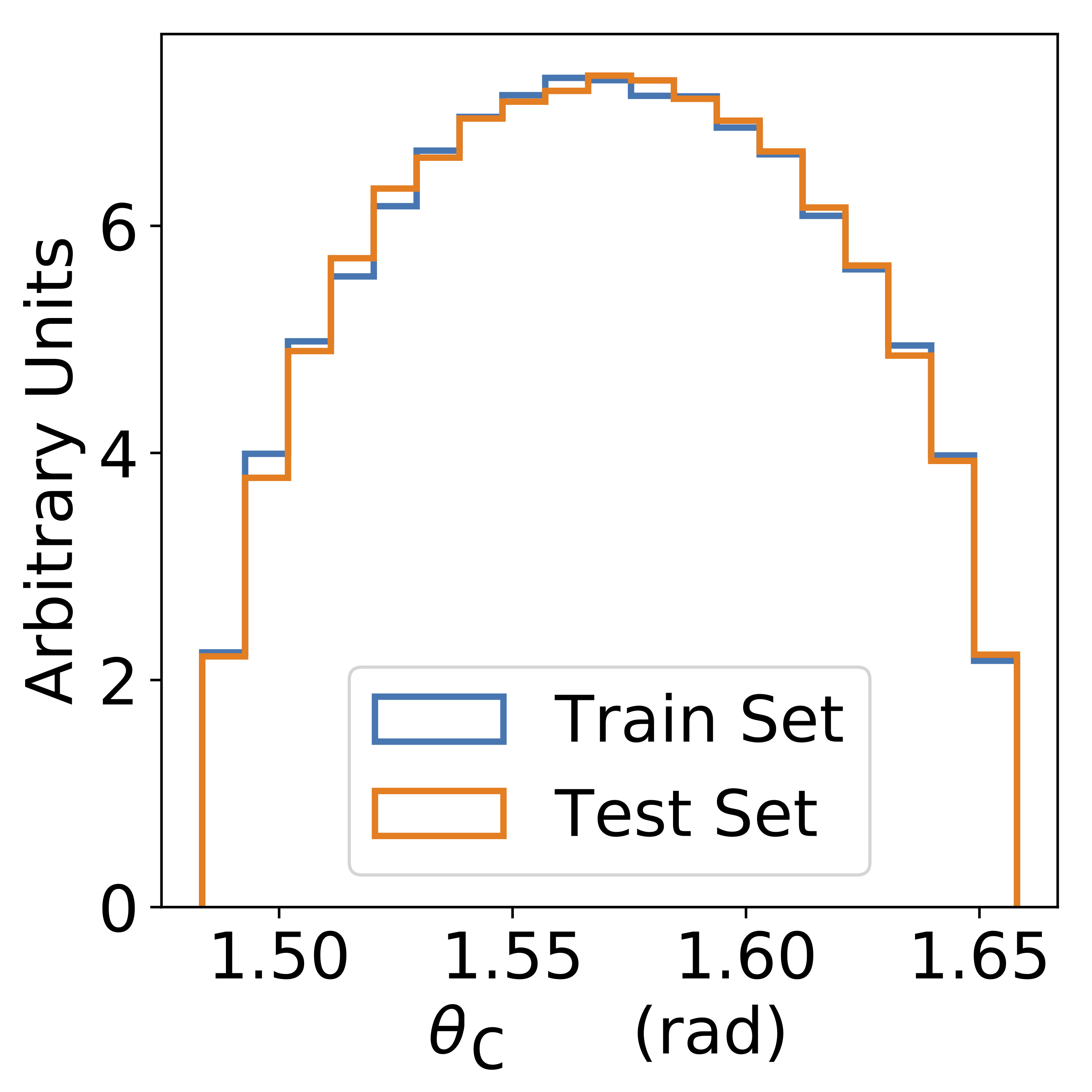}
}
\subfigure[Displacement in the coarsely segmented direction]{
\centering
        \includegraphics[width=0.45\textwidth]{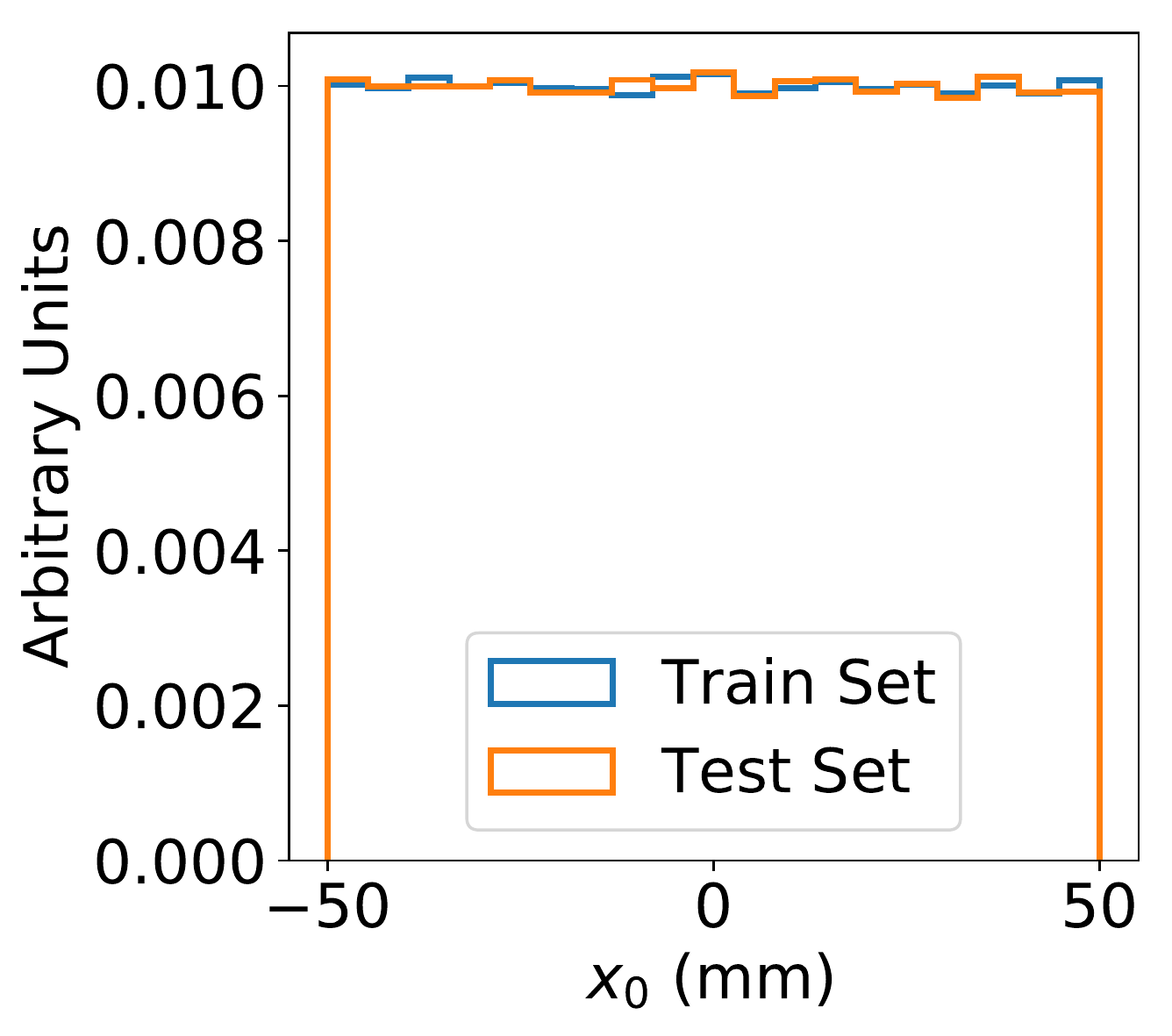}
}
\subfigure[Displacement in the finely segmented direction]{
\centering
        \includegraphics[width=0.45\textwidth]{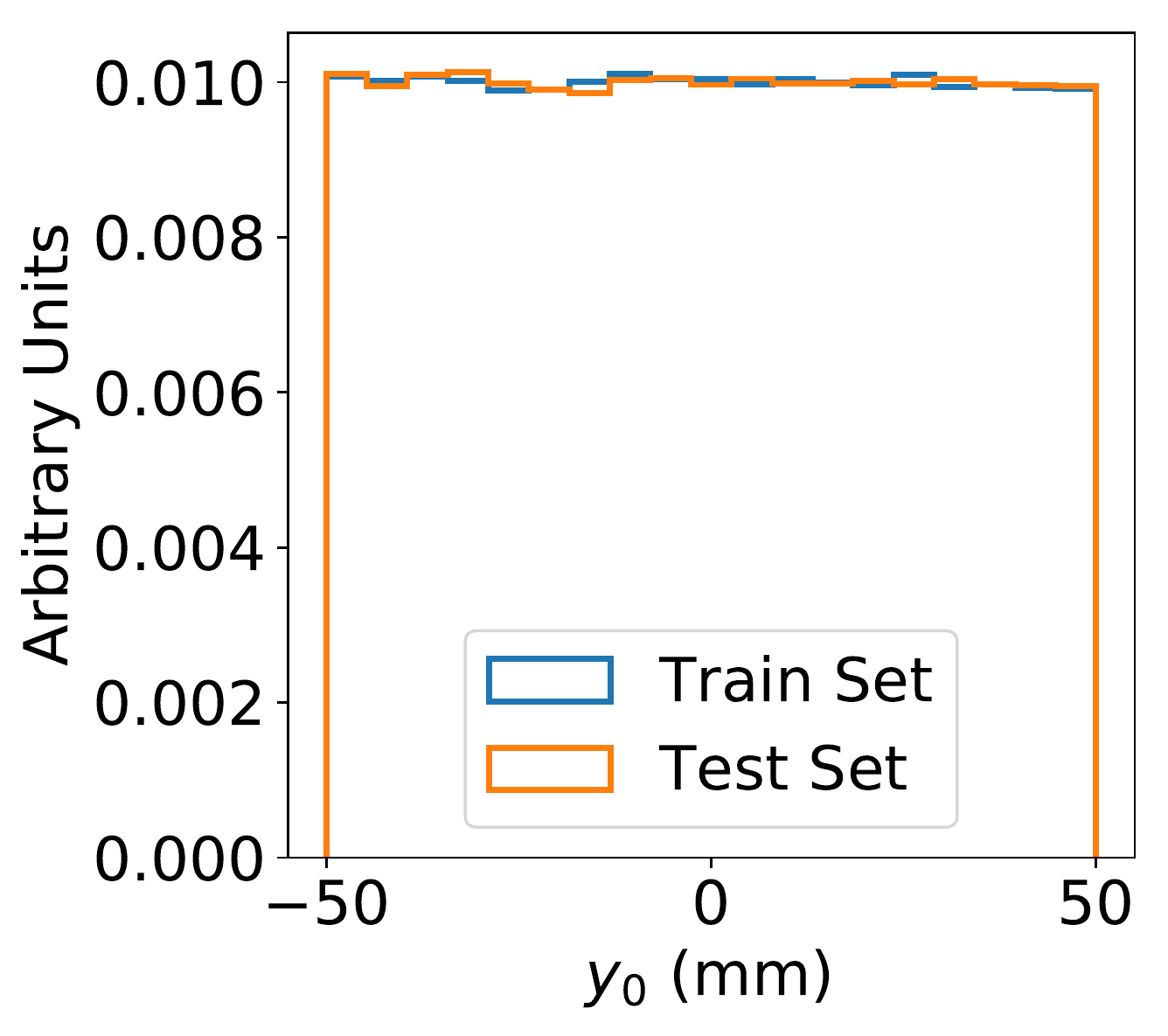}
}
\caption{Distribution of variations in the angle and position of incidence in the train and test sets for the positron data set. Other particle types also exhibits variations drawn from these distributions.}
\label{fig:distr}
\end{figure}

The incoming particles are shot isotropically in a circle around the $\hat{z}_{G}$ with a maximum aperture angle $\theta_{G}$ of $5^\circ$ (Fig.~\ref{fig:target}) and uniform lateral displacement of 5 cm in both $\hat{x}_{G}$ and $\hat{y}_{G}$ directions (Fig.~\ref{fig:distr}). Given the transformation in Eq.~\ref{eq:transform}, we obtain the following equations:
\begin{align}
    &p_x^G = p_y^C & p\ \mathrm{sin}\theta_G\ \mathrm{cos}\phi_G &=p\ \mathrm{sin}\theta_C\ \mathrm{sin}\phi_C \\
    &p_y^G = p_z^C & p\ \mathrm{sin}\theta_G\ \mathrm{sin}\phi_G &=p\ \mathrm{cos}\theta_C \\
    &p_z^G = p_x^C & p\ \mathrm{cos}\theta_G &=p\ \mathrm{sin}\theta_C\ \mathrm{cos}\phi_C
\end{align}
The distribution of angles in collider coordinates (Fig.~\ref{fig:distr}) can therefore be obtained from the momenta in \textsc{Geant4} coordinates:
\begin{align}
\theta_C &= \mathrm{cos}^{-1} \left( \frac{p_y^G}{p}\right)\\
\phi_C &= \mathrm{tan}^{-1} \left( \frac{p_x^G}{p_z^G}\right)
\end{align}

\end{document}